\documentclass[sigconf,sigconf]{acmart}
\AtBeginDocument{%
  }

\copyrightyear{2025}
\acmYear{2025}
\setcopyright{cc}
\setcctype{by}
\acmConference[CHI '25]{CHI Conference on Human Factors in Computing Systems}{April 26-May 1, 2025}{Yokohama, Japan}
\acmBooktitle{CHI Conference on Human Factors in Computing Systems (CHI '25), April 26-May 1, 2025, Yokohama, Japan}\acmDOI{10.1145/3706598.3713693}
\acmISBN{979-8-4007-1394-1/25/04}

\usepackage{algorithm}

\usepackage{multirow}

\newcommand{\bpstart}[1]{\vspace{1mm} \noindent{\textbf{#1}}.}

\newcommand{\name}[1]{\textsc{COALA}}
\newcommand{\add}[1]{\textcolor{black}{#1}}
\newcommand{\delete}[1]{}

\definecolor{WColor}{HTML}{BDBDBD}
\definecolor{WCColor}{HTML}{A6CEE3}
\definecolor{WEColor}{HTML}{1F78B4}
\definecolor{ASColor}{HTML}{B2DF8A}
\definecolor{PSColor}{HTML}{33A02C}
\definecolor{NTColor}{HTML}{FB9A99}

\setcopyright{cc}
\setcctype{by}
\copyrightyear{2025}
\acmYear{2025}
\acmDOI{10.1145/3706598.3713693}
\acmISBN{979-8-4007-1394-1/25/04}
\acmConference[CHI '25]{CHI Conference on Human Factors in Computing Systems}{April 26--May 01, 2025}{Yokohama, Japan}
\acmBooktitle{CHI Conference on Human Factors in Computing Systems (CHI '25), April 26--May 01, 2025, Yokohama, Japan}

\begin{document}
\title{Comparing Native and Non-native English Speakers' Behaviors in Collaborative Writing through Visual Analytics}

\author{Yuexi Chen}
\orcid{0000-0001-8720-5516}
\affiliation{\institution{Department of Computer Science \\ University of Maryland}
\city{College Park}
\state{Maryland}
\country{USA}}
\email{ychen151@umd.edu}

\author{Yimin Xiao}
\orcid{0000-0001-7482-5716}
\affiliation{\institution{College of Information \\ University of Maryland}
\city{College Park}
\state{Maryland}
\country{USA}}
\email{yxiao@umd.edu}

\author{Kazi Tasnim Zinat}
\orcid{0000-0001-7914-5955}
\affiliation{\institution{Department of Computer Science \\ University of Maryland, College Park}
\city{College Park}
\state{Maryland}
\country{USA}}
\email{kzintas@umd.edu}

\author{Naomi Yamashita}
\orcid{0000-0003-0643-6262}
\affiliation{\institution{NTT}
\city{Keihanna}
\country{Japan}}
\email{naomiy@acm.org}

\author{Ge Gao}
\orcid{0000-0003-2733-2681}
\affiliation{\institution{College of Information \\ University of Maryland}
\city{College Park}
\state{Maryland}
\country{USA}}
\email{gegao@umd.edu}

\author{Zhicheng Liu}
\orcid{0000-0002-1015-2759}
\affiliation{\institution{University of Maryland}
\city{College Park}
\state{Maryland}
\country{USA}}
\email{leozcliu@umd.edu}

\begin{abstract}
Understanding collaborative writing dynamics between native speakers (NS) and non-native speakers (NNS) is critical for enhancing collaboration quality and team inclusivity. In this paper, we partnered with communication researchers to develop visual analytics solutions for comparing NS and NNS behaviors in 162 writing sessions across 27 teams. The primary challenges in analyzing writing behaviors are data complexity and the uncertainties introduced by automated methods. In response, we present \textsc{COALA}, a novel visual analytics tool that improves model interpretability by displaying uncertainties in author clusters, generating behavior summaries using large language models, and visualizing writing-related actions at multiple granularities. We validated the effectiveness of \textsc{COALA} through user studies with domain experts (N=2+2) and researchers with relevant experience (N=8). We present the insights discovered by participants using \textsc{COALA}, suggest features for future AI-assisted collaborative writing tools, and discuss the broader implications for analyzing collaborative processes beyond writing.
\end{abstract}

\begin{CCSXML}
<ccs2012>
  <concept>
        <concept_id>10003120.10003145.10003151</concept_id>
        <concept_desc>Human-centered computing~Visualization systems and tools</concept_desc>
        <concept_significance>500</concept_significance>
    </concept>
   <concept>
        <concept_id>10003120.10003130.10011762</concept_id>
        <concept_desc>Human-centered computing~Empirical studies in collaborative and social computing</concept_desc>
        <concept_significance>500</concept_significance>
    </concept>
    <concept>
        <concept_id>10003120.10003145</concept_id>
        <concept_desc>Human-centered computing~Visualization</concept_desc>
        <concept_significance>500</concept_significance>
    </concept>
</ccs2012>
\end{CCSXML}

\ccsdesc[500]{Human-centered computing~Interactive systems and tools}
\ccsdesc[500]{Human-centered computing~Collaborative and social computing}
\ccsdesc[500]{Human-centered computing~Visualization}

\keywords{Collaborative writing, non-native speakers, event sequence analysis, \add{human-AI interaction}, \add{collaboration}, \add{large language models}}

\maketitle

%%%% 1-introduction.tex starts here %%%%

\section{Introduction}

Non-native speakers (NNS) actively engage in collaborative writing across various contexts: international students write reports with classmates~\cite{cheng2013non} and advisors~\cite{yang2021supervisor}; Wikipedia contributors edit articles in different languages~\cite{hale2014multilinguals}; employees in multi-national corporations collaborate on project pages~\cite{confluence}. Previous research shows that when writing in a non-native language, NNS tend to produce shorter and less complex content compare to writing in their native language~\cite{chenoweth2001fluency, wolfersberger2003l1, uzawa1989writing}. However, limited research has examined the collaborative writing processes involving NNS and how they write differently from native speakers (NS)~\cite{xiao2024dis}.
%the dynamics between NNS and native speakers(NS)~\cite{xiao2024dis}. 
Such knowledge will be insightful in enhancing collaboration outcomes and fostering inclusive team dynamics involving NNS. 

To bridge this gap, in collaboration with communication researchers, we collected 
% Therefore, we collaborate with two communication researchers, who have collected 
document history and screen recordings of 162 collaborative writing sessions from 27 teams. We are interested in \textit{comparing} authors' behaviors across different linguistic backgrounds and writing stages. 
% \leo{add a sentence or two to explain what we are trying to compare. This is an important thing we should elaborate on.}
For example, what are the common behavior patterns of NS and NNS, respectively? How do they differ in the early and late stages of collaborative writing?

% Given history versions of documents, text
However, existing document visualization and analytics tools fall short of supporting our analysis goal. Visualizations of document versions have long been used to investigate the dynamics of collaborative writing. For example, History Flow~\cite{viegas2004studying} uses a Sankey diagram to encode content contributions from different Wikipedia authors across different versions; Time Curves~\cite{bach2015time} project different document versions on a 2D curve based on their similarity and temporal order. However, such visualizations focus on the document content instead of the writers' behaviors during the writing process, such as browsing the internet or using translation tools. 

% Authors' behaviors could be treated as event sequences, and visual analytics have been used to compare and analyze sequence patterns. 
Event sequence visualization tools are better suited for behavioral data but still do not effectively meet our needs. 
For example, TipoVis~\cite{han2015visual} 
% helps psychology researchers compare action sequences representing social and communicative behaviors of children,  but 
allows comparisons between two sequences at a time, while our analysis requires comparing behavior sequences across multiple authors and sessions. CoCo~\cite{malik2015cohort} supports comparing event sequences belonging to different cohorts, but focuses primarily on aggregated metrics such as frequencies. In our case, we need to identify behavioral differences between NS and NNS at multiple levels of granularity with meaningful qualitative descriptions. 
% in terms of actions taken by doctors. However, existing event sequence analytics are not directly applicable in our case. For instance, TipoVis~\cite{han2015visual} only supports comparing two sequences at a time, and CoCo returns aggregated metrics by cohorts, ignoring individual sequences.

Furthermore, to effectively analyze such complex behavioral datasets, it is important to combine visualizations with automated methods such as data mining and clustering. Previous research highlights challenges related to the interpretability and trustworthiness of automated methods
% visualization alone does not address our research questions. Since our dataset has 27 teams and 162 writing sessions, automated methods are required to group and summarize sequences. 
% However, understanding and trusting of automatic methods' output can not be taken for granted
~\cite{chuang2012interpretation, chatzimparmpas2020state, liang2023multiviz, yang2020visual}. 
% \leo{cite more papers, Jason's paper is a little dated, there are more recent ones.} \tracy{added} 
As communication research experts, our collaborators 
% communication researchers 
possess contextual knowledge about their dataset and multilingual communication in general but are not familiar with the mechanisms of automated methods. 
Before incorporating the model-generated results into their analysis, they must interpret and trust them.

In comparing the behaviors of NS and NNS in collaborative writing, we thus face two challenges: 1) the limitations of existing text and event sequence visual analytics approaches, and 2) the lack of interpretability and trust in automatic data analytics. 
% \leo{it is weird to jump directly from challenges to contributions, describe our approach/method to address the challenges before talking about contributions. For challenge 1, what did we do? For challenge 2, what did we do?} \tracy{(It's a bit repetitive if I talk about both solution and contributions)} \leo{I'm not saying we talk about solutions here, we talks about METHODS.} 
To address these challenges, we worked closely with the communication experts to formulate data models and task requirements, iterated on visualization designs to assess their applicability and scalability, and identified factors that might hinder interpretation and trust. Based on these design iterations, we make the following contributions: 

\begin{itemize}
    \item \name{}, a visual analytics tool for comparing native and non-native speakers' behaviors in collaborative writing. \name{} displays the uncertainties of multiple clustering results and supports interactive refinement of clusters while leveraging large language models (LLM) to generate cluster summaries. 
    
    \item Empirical validation of \name{} through a focus group session (N=\add{2+2}) and \delete{2) a comparison with an established sequence visualization tool (CoCo \cite{malik2015cohort}) in terms of their support for behavior clustering and summarization.}\add{ individual study sessions with researchers in related fields (N=8)}, where the participants used \name{} to analyze behavioral differences between NS and NNS.  
    %\zinat{Should we just write `user study' instead of `an individual user study'?}
    
    \item Design lessons for developing interpretable visual analytics in the context of communication research and findings that inform future AI-assisted collaborative writing tools \add{and collaborative processes beyond writing.}   
\end{itemize}

%%%% 1-introduction.tex ends here %%%%

%%%% 2-related-work.tex starts here %%%%

\section{Related Work}\label{relatedWork}

\subsection{Collaborative writing studies}

Collaborative writing has been a topic of interest since the 1980s~\cite{engelbart1984collaboration, fish1988quilt, baecker1993user}. Early research focused on awareness and coordination during collaborative writing~\cite{dourish1992awareness}, common writing tasks, and the number of collaborators~\cite{kim2001reviewing}. The rise of online collaborative writing tools like Google Docs, Microsoft Word, and Overleaf~\cite{overleaf} has made collaborative writing a common practice. For instance, Olson et al.~\cite{olson2017people} analyzed collaborative writing patterns in 96 college assignments in Google Docs and found that balanced participation and leadership would result in higher writing quality~\cite{olson2017people}. Researchers also explored various aspects of collaboration, such as impression management~\cite{birnholtz2013write}, reluctance to write closely~\cite{wang2017users},
preference over edits with explanations~\cite{park2023importance}, differences of tasks across writing stages~\cite{sarrafzadeh2021characterizing}, and territorial behaviors~\cite{larsen2019territorial}.

Few studies focus on authors' off-document writing-related activities during collaborative writing, for example, navigating multiple applications (e.g., Google Docs and Adobe InDesign)~\cite{larsen2020collaborative} and coordinating writing tasks on Wikipedia discussion pages~\cite{schneider2011understanding}. 
Collaborated with communication researchers, we focus on writing-related behaviors, including off-document behaviors like using a translator and browsing the internet, aiming to provide a new perspective on collaborative writing analysis. 

%Olson et al.~\cite{olson2017people} analyzed collaborative writing patterns in 96 college class assignments on Google Docs and found that balanced participation and the presence of leadership will result in higher writing quality. Birnholtz et al.~\cite{birnholtz2013write, birnholtz2012tracking} studied group maintenance, impression management, and relationship-focused behaviors. Park et al.~\cite{park2023importance} conducted a collaborative writing experiment of 20 pairs of authors and found that participants prefer co-writers who provide rationales for edits. Sarrafzadeh et al. ~\cite{sarrafzadeh2021characterizing} analyzed user interaction logs (e.g., interacting with paragraphs, find-and-replace, etc.) on a collaborative writing platform at different temporal stages. Larsen-Ledet et al. studied territorial functioning in collaborative writing and emphasized the importance of understanding interpersonal dynamics and the temporal aspects of writing, such as timing and turn-taking. Larsen-Ledet et al.~\cite{larsen2019territorial} studied how territorial functioning in collaborative writing, and emphasized the importance of understanding interpersonal dynamics and the temporal aspects of writing, such as timing and turn-taking.

\subsection{Text visual analytics}
Several visual analytics approaches have been designed to analyze the evolution of documents in collaborative writing. Itero~\cite{turkay2018itero} is a revision history analytics tool based on Google Docs that visualizes character insertion patterns and user contributions. History Flow~\cite{viegas2004studying} and DocuViz~\cite{wang2015docuviz} encode each author's contribution as a colored vertical line, with the height of the line proportional to the content length. The flow-like visualization reveals the cooperation and conflict among co-authors by connecting the same line across different versions. Graphs are also widely used, where authors are represented as nodes, and edges could be disagreement~\cite{flock2015whovis} or revert actions~\cite{kittur2007he}. Time Curves~\cite{bach2015time} is a timeline visualization based on points' similarity, which could visualize different document versions. Other visualizations include branch-based visualizations~\cite{perez2018organic}, revision maps~\cite{southavilay2013analysis} and color-coded words by authorship~\cite{flock2015towards, torres2019visualizing}.
Compared to text visual analytics approaches, \name{} focuses on sequences of writing-related behaviors.

\subsection{Non-native speakers vs. native speakers}
Compared to native speakers (NS), non-native speakers (NNS) usually produce shorter and less complicated content and have difficulty transferring writing strategies from their mother tongue~\cite{chenoweth2001fluency, wolfersberger2003l1, uzawa1989writing}. Though NNS need more help in the expression aspects~\cite{severino2009comparison}, they may still contribute to the ideation aspect~\cite{goffman1981forms}. Cheng et al.~\cite{cheng2013non} found in a case study that NS students had more power in collaborative writing at the beginning, but the NNS student developed academic literacy along the way, and overall the group writing experience has improved. NNS' writing could also be improved by receiving direct edit feedback at early versions~\cite{karim2020revision, yang2021supervisor}, or exposure to well-written model text by NS~\cite{kang2020using}. Compared to these previous case studies, we have a larger collaborative writing dataset of NS-NNS with video-recorded author behaviors, poised to reveal more patterns beyond anecdotal evidence.

\subsection{Event sequence analysis}
There are numerous methods to analyze event sequences. Besides \textit{visualizing} sequences, we categorize analysis methods based on tasks: \textit{comparing}, \textit{clustering} and \textit{summarizing}.

\bpstart{Visualizing event sequences}
The most straightforward visualization design for event sequences is to arrange the events on a timeline~\cite{plaisant1996lifelines, krstajic2011cloudlines}. When the number of sequences is large, flow-based visualizations could show the trend of bundled sequences. For example, Sankey diagrams represent each event as a node, the length of the node and the thickness of the links between nodes encode event frequencies~\cite{wongsuphasawat2011outflow, perer2014frequence}. Tree-based visualizations encode the frequency of events as the thickness of edges~\cite{hu2016visualizing, liu2017coreflow}. Like tree-based visualizations, icicle plots encode events as stacked rectangles, ordered from top to bottom, usually colored by event categories~\cite{wongsuphasawat2011lifeflow, liu2016patterns}. When subsequences are highly repetitive, matrix-based visualizations could show the transition trend clearly~\cite{perer2012matrixflow, zhao2015matrixwave}.

\bpstart{Comparing event sequences}
Multiple tools focus on comparison. CoCo compares two patient cohorts via statistical analysis with built-in metrics~\cite{malik2015cohort} distilled from domain expertise~\cite{monroe2012exploring}. TipoVis compares event sequences of social and communicative behaviors by overlaying two sequences~\cite{han2015visual}. COQUITO~\cite{krause2015supporting} assists users in defining cohorts with temporal constraints and comparing sequences by overlapped branches.
Directly linking event sub-sequences for comparison is also common~\cite{qi2019stbins, meyer2009mizbee, zhao2023contextwing}.  

\bpstart{Clustering event sequences}
Several interactive tools are designed for clustering sequences. For example, Wang et al.~\cite{wang2016unsupervised} built an unsupervised interactive clustering system to analyze large-scale clickstream data. EventThread~\cite{guo2017eventthread} clusters event sequences by latent stage categories. Gotz et al.~\cite{gotz2019visual} group event sequences by dynamic hierarchical dimension aggregation. Sequen-C~\cite{magallanes2021sequen} adopts an align-score-simplify strategy to cluster sequences. VASABI~\cite{nguyen2019vasabi} clusters user profiles by topic modeling and uses multi-dimensional distributions to characterize each cluster.

\bpstart{Summarizing event sequences}
Numerous methods have been developed to find an overview of a cluster of sequences. Sequence Synopsis~\cite{chen2017sequence} constructs a high-level overview of sequences by balancing the minimum description length (MDL) principle and the information loss. CoreFlow~\cite{liu2017coreflow} extracts branching patterns in temporal event sequences. Frequence~\cite{perer2014frequence} is a visual analytics tool built on a frequent pattern mining algorithm that handles multiple levels of details and concurrency. SentenTree~\cite{hu2016visualizing} summarizes unstructured social media text in a tree structure. 

\name{} is also equipped with visualizing, comparing, clustering, and summarizing features. Compared to existing approaches, we focus on interpretation and trust by including multiple clustering methods and displaying uncertainties, supporting multi-level granularity of sequences, and leveraging large language models to generate more intuitive descriptions~\cite{brown2020language}.

%%%% 2-related-work.tex ends here %%%%

%%%% 3-background.tex starts here %%%%

\section{Study Background}\label{domain-background}
%\leo{I suggest we do not divide the content into subsections. Instead, we just provide a concise description of the collaborators, the design study process, and the motivation to collect and analyze data. We do not need statistics on non-native speakers in sec 2.2.}\tracy{merged}
\subsection{Background}
We collaborated with two communication researchers from a public university in the US. One is a professor who has studied multilingual communication for more than a decade, and the other is a Ph.D. advisee of the professor who also has rich experience in multilingual communication. They collected a dataset of collaborative writing between native and non-native speakers and contacted us for suggestions in visual analytics. 

We conducted longitudinal \delete{interviews}\add{co-design sessions} with our collaborators to understand the communication research analysis better. We met weekly or bi-weekly for 30 weeks. After they introduced the study background and data, we initially adapted existing text visualizations like Time Curves~\cite{bach2015time} and History Flow~\cite{viegas2004studying} (see Appendix). \add{Though such text visualizations provide a glimpse of how authors contribute to the document, and how co-authors revise or delete each others' contribution,}\delete{However,} \delete{such text visualizations}\add{they} do not address the research questions to compare authors' behaviors, so we designed dedicated features for analyzing authors' behavioral sequences. During the process, we showed them visualizations and interface mockups, incorporated the feedback into the next iteration, and finalized the visualization and interaction designs with them. 
\subsection{Data collection}
\begin{figure}
    \centering
    \includegraphics[width=0.5\textwidth]{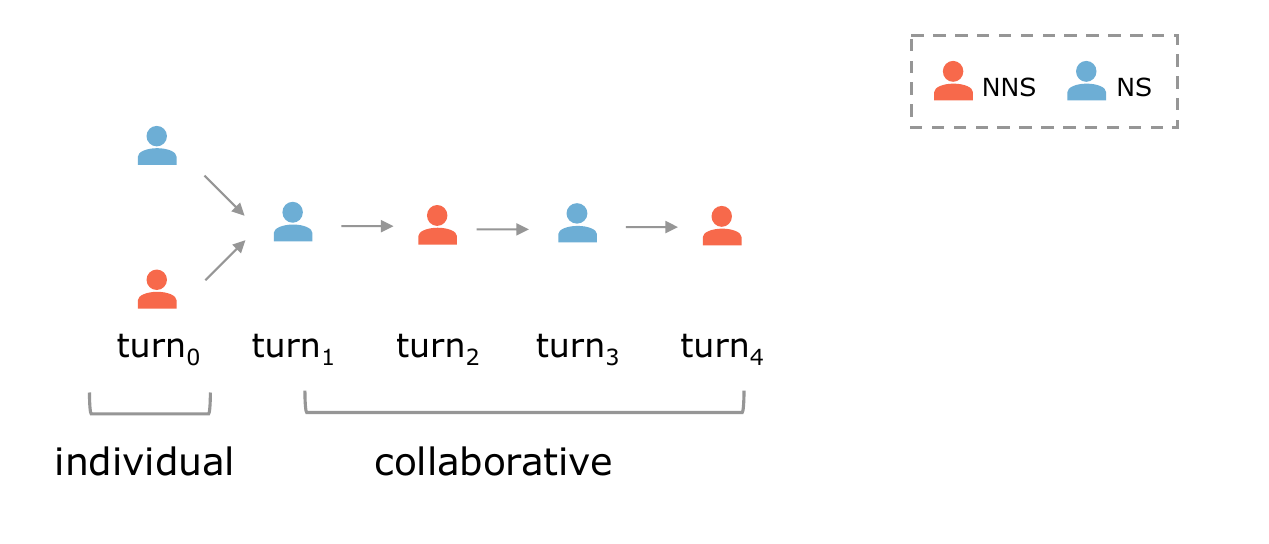}
    \caption{Turn-taking of NS (native speaker) and NNS (non-native speaker) of English in the multilingual collaborative writing study.
    % \leo{reduce the white space in the figure, put a border around the legend, or annotate the icons directly}\tracy{revised}
    }
    \label{fig:turn-taking}
    \Description{A visualization of turn-taking between native speakers (NS) and non-native speakers (NNS) in a multilingual collaborative writing study. Initially, in turn 0, both NS and NNS write independently. In turns 1 to 4, they take turns contributing to the same document.}
\end{figure}
Our collaborators recruited 29 native English speakers (NS) and 29 non-native English speakers (NNS) from an American university
%(N = 29, 27 F, 1 M, 1 non-binary) 
and a Japanese university 
%(N = 29, 16 F, 13 M) 
%Following the previous experiment setup on collaborative writing~\cite{barile2002computer, birnholtz2013write, galegher1992computer},
for an online collaborative writing study. \add{To ensure NNS are of similar English proficiency,} all NNS are native Japanese speakers with \textit{limited working proficiency} in English~\cite{ILRscale69:online}. Participants are asked to act as if they were columnists for an English magazine who answer readers' questions about the role of technology in modern life. The topics include social media, remote learning, and digital privacy. \add{To mitigate the influence of topic familiarity, } one NS and one NNS are paired to form a writing group and are assigned a topic familiar to both authors. All participants are provided with preset Google accounts and links to blank Google Docs. NNS and NS of the same group do not know each other but are informed about co-authors' language proficiency. Participants are also asked to record their screen during writing. \add{Participants are welcome to use any tools (e.g., search engines, Google Translate, Grammarly) that they normally use during writing. Since the study was conducted before ChatGPT's release, no participant used generative AI tools.}

Then, NNS and NS take turns writing an English essay jointly. The turn-taking setup is shown in Figure~\ref{fig:turn-taking}: first, each author writes independently ({turn$_0$}), ensuring they have actively thought about the task instead of being a free-rider. Then, NS review the write-ups of both authors from {turn$_0$} and merges them into a single document ({turn$_1$}). During {turn$_1$}, NS can add/delete/edit any content. Next, NNS revise the joint document {turn$_2$}, followed by NS {turn$_3$}, and finally concluded by NNS {turn$_4$}. \add{Participants are allowed up to 1 hour for each turn, and they could finish early if they have nothing more to contribute.} After removing two teams that did not follow instructions, we have 27 remaining. 

% \leo{we need to explain the rationale of choosing an experiment setup instead of collecting data from a more ecologically valid setting like Wikipedia datasets. We also need to explain the rationale of choosing this particular experiment design (turn taking).}

Our collaborators carefully designed the setup of the above experiment. First, though current online writing tools support simultaneous writing, turn-taking is still a popular workflow adopted by co-writers in practice, as they minimize the burden of syncing content mentally~\cite{boellstorff2013words}, protect authors' territoriality~\cite{larsen2019territorial} and thus promote editing other co-authors' writing~\cite{andre2014effects}. Second, NNS may face difficulties identifying opportunities for contribution in flexibly structured collaboration with NS. Previous research has introduced several techniques to interrupt the natural conversation flow and impose contribution opportunities of NNS, including artificial silent gaps~\cite{yamashita2013lost} and a conversational agent~\cite{li2023improving}. Therefore, having a designated opportunity to ensure NNS contributes to collaborative writing is necessary. Besides, since it's one of the first studies on multilingual collaborative writing, researchers chose a simplified setup \delete{and only included}\add{with only} one NS and one NNS co-writer in a team to pinpoint the \add{group} dynamics\delete{between co-writers} easily, leaving more complex configurations for future research \add{(e.g., NNS from different countries, unbalanced group settings where there are more NS than NNS or vice versa)}.

%%%% 3-background.tex ends here %%%%

%%%% 4-data.tex starts here %%%%

\section{Data Abstraction}\label{data-abstract}
% \leo{consider moving 3.4 and 3.5 to a section called Dataset Description and Data Model. We should have a more high-level abstraction of the data model beyond the current description that is semantically rich and domain-specific.}
%\subsection{Data abstraction}\label{data-abstract}
%\leo{Merge sec 3.2 and 3.3. Let's call the merged subsection Data Abstraction.}\tracy{merged}

Based on previous research, we introduce a general data model for the collaborative writing processes in our study.

\bpstart{Authors} 
In collaborative writing, there must be at least two authors. Let $A = \{a_1, a_2, ...\}$ be the set of authors. For each author $a_i$, the meta-information $M_i$ is a set of the author's quantitative or qualitative attributes, e.g., author's ID, linguistic background, seniority and so on. In our study, there are two authors in a team, so $A = \{a_{1}, a_{2}\}$. Since we only care about the linguistic background of authors in this study, $M_1 = $ \{\textit{native-English-speaker}\} and $M_2 =$ \{\textit{non-native-English-speaker}\}.

%One example is that Wikipedia authors have 23 different editor levels based on their accumulated contribution, from ``Registered Editor'' to ``Sagacious Editor''~\cite{editorLevel}. \leo{we are not analyzing the wikipedia dataset in this paper, better not mention it.}

\bpstart{Events} Let $E$ be the set of all collaborative writing events. The events could be on-document ($E_{on}$) and related ($E_{related}$), and $E = E_{on} \cup E_{related}$. Event $e_{on,i}$ is an edit made by authors, e.g., $e_{on,i}$ could include the author, editing types (add/delete), locations, and so on. Event $e_{related,i}$ is an event related to the collaborative writing process, but not limited to editing, e.g., leave a comment~\cite{zhang2019modeling}, make a post~\cite{schneider2011understanding} and browse the internet. $e_{related,i}$ could include the event name, start and end time. 
%\leo{from the perspective of data collection, the distinction between on-document vs. related events make sense, but why do we care from the perspective of visual analytics?}\tracy{one is about content, one is about behaviors?} 
We can automatically obtain $E_{on}$ by comparing different document versions. To obtain $E_{related}$, one communication researcher manually coded events by watching the recordings, including the author, event type, and start and end time. After that, another communication researcher helped categorize events into higher levels. There are six types of events in total, and we highlight each event in different colors: 

\noindent\colorbox{WColor}{\small \textbf{Writing}}: activities include ``On Google Docs'', ``Using online editing tools'', ``Checking for creating citation''.

\noindent\colorbox{NTColor}{\small \textbf{Note-taking}}: activities include ``Writing a note''. The difference between ``Writing'' and ``Note-taking'' is that the note does not go into the writing, but serves as an auxiliary role.

\noindent\colorbox{WCColor}{\small \textbf{Wordsmith-crosslingual}}: though the goal is to write an English essay, many NNS chose to write in Japanese and translate to English, or translate the write-up and read in Japanese. Such activities include ``Using translators to read'', ``Using translators to write'', ``Searching for language-related information'', ``Checking a dictionary or thesaurus'' and ``Checking for meanings''.

\noindent\colorbox{WEColor}{\small \textbf{Wordsmith-English}}: some NS also seek help with expression, such activities include ``Checking dictionary or thesaurus'', ``Searching for language-related information'' and ``Checking for meanings''.

\noindent\colorbox{ASColor}{\small \textbf{Active-search}}: authors may seek external evidence to build their argument. Activities include ``Searching for online information''.

\noindent\colorbox{PSColor}{\small \small \textbf{Passive-search}}: After an author cites an external source in the collaborative write-up, the other author may check the content. Such activities include ``Opening a URL to read information''.

Table~\ref{tab:videoActivityFreq} shows summarized frequencies and duration of six high-level writing-related actions: ``Writing'' is the most frequent and time-consuming action, followed by ``Wordsmith-crosslingual'', ``Active-search'', ``Passive-search'', ``Wordsmith-English'' and ``Note-taking''.

%\begin{table}[htbp]
%    \centering
%    \begin{tabular}{lll}
%    \toprule
%         Writing-related actions & Duration (hours) & Frequency \\
%    \midrule
%        \colorbox{WColor}{\small Writing} & 91.7 (73.3\%) & 2254 (52.3\%)  \\
%         \colorbox{WCColor}{\small Wordsmith-crosslingual}& 18.8 (15.0\%) & %1281 (29.7\%) \\
%         \colorbox{ASColor}{\small Active-search}& 13.2 (10.5\%) & 658 %(15.3\%) \\
%         \colorbox{PSColor}{\small \small Passive-search}& 0.8 (0.7\%) & 69 (1.6\%) \\
%         \colorbox{WEColor}{\small Wordsmith-English}& 0.5 (0.4\%) & 46 (1.1\%)  \\
%         \colorbox{NTColor}{\small Note-taking} & 0.1 (0.1\%) & 4 (0.1\%)\\
%    \bottomrule
%    \end{tabular}
%    \caption{Duration and frequencies of writing-related actions from video recordings}
    %\label{tab:videoActivityFreq}
%\end{table}

\begin{table*}[h]
    \centering
    \begin{tabular}{lcccc}
        \hline
        \multirow{2}{*}{Action} & \multicolumn{2}{c}{Duration} & \multicolumn{2}{c}{Frequency} \\
        & \add{NS} & \add{NNS} & \add{NS} & \add{NNS} \\
        \hline
        \colorbox{WColor}{\small Writing} & 52.5h (84.9\%) & 39.3h (61.9\%) & 706 (57.5\%) & 1548 (50.2\%) \\
        \colorbox{PSColor}{\small \small Passive-search} & 0.7h (1.1\%) & 0.1h (0.2\%) & 47 (3.8\%) & 22 (0.7\%) \\
        \colorbox{ASColor}{\small Active-search} & 8.1h (13.1\%) & 5.1h (8.1\%) & 425 (34.6\%) & 233 (7.6\%) \\
        \colorbox{WEColor}{\small Wordsmith-English} & 0.5h (0.8\%) & 0.0h (0.0\%) & 45 (3.7\%) & 1 (0.0\%) \\
        \colorbox{WCColor}{\small Wordsmith-crosslingual} & 0.1h (0.1\%) & 18.7h (29.6\%) & 4 (0.3\%) & 1277 (41.4\%) \\
        \colorbox{NTColor}{\small Note-taking} & 0.0h (0.0\%) & 0.1h (0.2\%) &     0 (0.0\%) & 4 (0.1\%)\\
        \hline
    \end{tabular}
    \caption{\add{Duration and frequencies of writing-related actions from video recordings}}
    \label{tab:videoActivityFreq}
\end{table*}

\bpstart{Turns and stages} 
% We can further group consecutive \textit{events} made by a single \textit{author} into a \textit{turn}. Let $T$ be all the turns, $T = \{t_1, t_2, ... \}$. Each turn $t_i = \{e_1, e_2, ... \}$. 
In collaborative writing, authors take turns to write, which could be either sequential or parallel~\cite{olson2017people}, depending on whether the timestamps of events overlap. In our study, as shown in Figure~\ref{fig:turn-taking}, $t_{0,ns}$ and ${t_{0,nns}}$ are turns where authors write individually, and $t_1$ to $t_4$ are collaborative turns. These turns are organized into writing stages; therefore, we have individual stages $t_{0,ns}$ for NS and $t_{0,nns}$ for NNS, and collaborative stages $\{t_{1,ns}, t_{3,ns}\}$ for NS and $\{t_{2,nns}, t_{4,nns}\}$ for NNS.

\bpstart{Document versions}
Let $V$ be the entire history versions of a single document, $V = \{v_1, v_2, ...\}$. Document $v_{i+1}$ is the result of event $E_i$ on $v_i$. Google Docs record word-level document history; however, our collaborators are not interested in such fine-grained analysis. Instead, they collected the document version at the end of each turn for each author $V_{end} = 
\{v_{0,ns}, v_{0,nns}, v_1, v_2, v_3, v_4\}$, then sampled three intermediate versions that reflected a person's writing progress during turn 2 to 4: \[
V_{inter} = \{ v_{i,j} \mid i \in \{2,3,4\},\; j \in \{1,2,3\} \}.
\] In total, we have 15 versions: $V = V_{end} \cup V_{inter}$.

% \tracy{revised sequences}
\bpstart{Sequences} The events happening in different turns from an author are grouped into a sequence based on the writing stages. As shown in Figure~\ref{fig:turn-taking}, we have four collections of sequences: $S_{NS, individual}$ are events in $turn_{0}$ by NS; $S_{NNS, individual}$ are events in $turn_{0}$ by NNS; $S_{NS, collaborative}$ are events in $turn_{1}$ and $turn_{3}$ concatenated; $S_{NNS, collaborative}$ are events in $turn_{2}$ and $turn_{4}$ concatenated. We chose to concatenate events in collaborative turns as suggested by communication researchers, as based on their experience, behaviors are similar across turns in the collaborative stage.

% \leo{What does stage mean? define it. Also label it in Fig 2.}\tracy{labeled}. \leo{it's better to talk about this later along Fig 3 after the tasks are introduced?}\tracy{talked in the task analysis sections}

%\begin{figure*}
%    \centering
%    \includegraphics[width=\textwidth]{figs/rawSequences.png}
%    \caption{Illustrations of event sequences of authors' behaviors grouped by writing stages \leo{I don't think we need this figure}}
%    \label{fig:raw-event-seq}
%\end{figure*}

%%%% 4-data.tex ends here %%%%

%%%% 5-methods.tex starts here %%%%

\section{Methods}\label{method-section}

%\section{Task Analysis}
\begin{figure*}
    \centering
    \includegraphics[width=\textwidth]{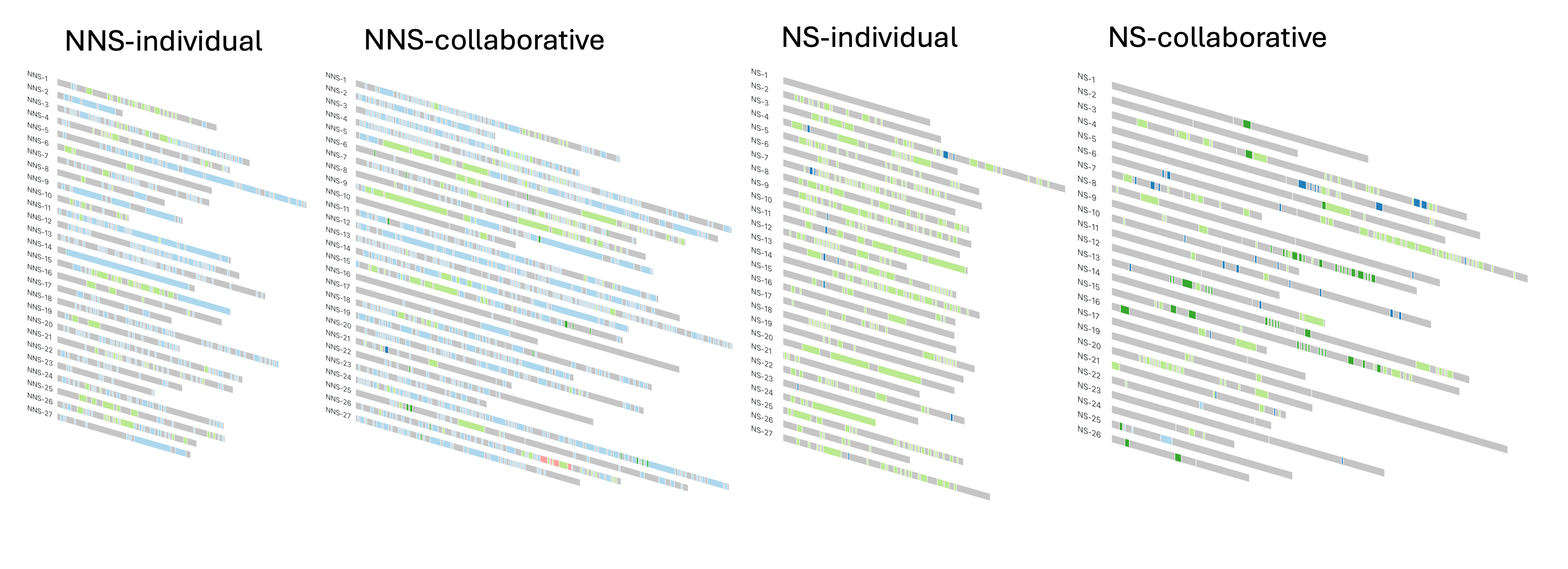}
    \caption{All sequences of non-native speakers' (NNS) and \add{native speakers's (NS)} behaviors during the individual \delete{(left)}and collaborative \delete{(right)} writing stage. The sequences in the collaborative writing stage are longer due to the concatenation of turns. The length of the rectangles indicates the duration of each event, and the color encodes the event types.} 
    \label{fig:all-events}
    \Description{A sequence visualization of NS and NNS behaviors during individual and collaborative writing. Rectangles represent events, with their lengths indicating event duration. The sequences vary in type and length.}
\end{figure*}

Communication researchers are interested in comparing authors' writing behaviors along two dimensions: writing stages (individual, collaborative), and author types (NS, NNS). For example, to answer the question ``during the collaborative writing stage, how do NS and NNS writers' behaviors differ?'', we need to compare two collections of sequences: $S_{NS, collaborative}$ and $S_{NNS, collaborative}$. 

A primary challenge is that the sequences are highly heterogeneous, e.g., sequences in $S_{NNS, collaborative}$ range from 6 to 188 events, with an average of 75.6 and standard deviation of 47.0, making direct visual comparison impossible, as shown in Figure~\ref{fig:all-events}. 

Therefore, we identified the following lower-level tasks to enable effective comparison:

\begin{itemize}
    \item \textbf{T1: cluster similar sequences within each collection.} Besides automatic clustering methods, communication researchers should be able to incorporate their domain expertise into the clustering results.

    \item \textbf{T2: summarize each sequence cluster within each collection.} The summarization should provide rationales to help communication researchers interpret clusters.
    
\end{itemize}

In this section, we discuss computational methods to support task \textbf{T1} and \textbf{T2}, the interpretation challenges we faced in the design study, and our strategies to solve those challenges. 

\subsection{Computational methods}
For T1, we decided to first automatically cluster sequences. Clustering algorithms are widely used to organize similar data points into groups~\cite{kaufman2009finding}. Common methods include k-means~\cite{hartigan1979algorithm}, hierarchical clustering~\cite{murtagh2012algorithms}, DBSCAN~\cite{schubert2017dbscan}, self-organizing maps~\cite{wei2012visual} and so on. For T2, we reviewed the literature on visual summarization and data mining in event sequence analytics~\cite{guo2021survey, zinat2023comparative, zinat2024multi}, and decided to use frequent patterns to summarize each cluster.

\subsubsection{Method I: cluster and summarize sequences jointly}
Our first approach is to cluster and summarize event sequences jointly. We chose Sequence Synopsis~\cite{chen2017sequence} because it is reported to produce higher quality visual summaries compared to other summarization methods~\cite{zinat2023comparative}. Besides, sequential patterns are the most widely used summarization format~\cite{perer2014frequence, polack2018chronodes, liu2016patterns, chen2017sequence}, compared to trees~\cite{liu2017coreflow} and directed acyclic graphs (DAG)~\cite{hu2016visualizing}. Sequence Synopsis clusters sequences and constructs a sequential pattern of each cluster by \delete{balancing the minimal description length and information loss}\add{striking a balance between pattern conciseness and minimizing information loss from the original sequences}. In our implementation, we can indirectly control the number of clusters $K$ and the length of patterns by adjusting the weights of information loss and the number of patterns.

\subsubsection{Method II: cluster first, summarize later}
Our alternative approach is to cluster sequences first, and summarize each cluster. Here, we considered k-means and hierarchical clustering because these algorithms allow us to easily change the number of clusters. For \delete{every two}\add{each pair of sequences}, we computed the Levenshtein distance (ignoring duration), \add{which returns the minimal number of edits required to align the two sequences. Compared to other common distance metrics such as Euclidean distance, Levenshtein distance captures the order of the sequence and handles sequences of different lengths.} Given $K$ clusters, we also prefer nested results. For example, if two sequences are in the same cluster when $K = 4$, then we expected them to still be in the same cluster when $K = 3$. Therefore, we chose hierarchical clustering over k-means~\cite{tan2016introduction}. Since hierarchical clustering algorithms do not generate visual summaries for each cluster, 
we ran a commonly used maximal pattern mining algorithm VMSP~\cite{fournier2014vmsp} to extract patterns for each cluster. 
We set the minimum support to be 50\%, i.e., the pattern has to be present in at least half of the sequences in the cluster. Different from Sequence Synopsis, which returns only one pattern for each cluster, VMSP returns multiple patterns, and we chose the longest one with the maximum support as the representative pattern. 

\add{Method I and method II differ in two key aspects: 1) the distance metric: method I does not rely on an explicit distance metric but learns how to cluster similar sequences and extract a representative pattern by balancing the pattern length and the information loss. In contrast, method II uses Levenshtein distance, which computes the edit distance between two sequences required for alignment; 2) the connection between patterns and clusters: for method I, the patterns and clusters are tightly coupled, each cluster is represented by a single pattern; for method II, clusters and patterns are loosely coupled, as pattern mining algorithms return multiple patterns for a cluster, it offers greater flexibility. By including both methods, \name{} explores the computational space more thoroughly, provides multiple approaches to the clustering and summarization tasks, and thus brings different perspectives to the datasets.}

\subsection{Challenges in interpreting the computational results}\label{interpretation-challenge}
\begin{figure}
    \centering
    \includegraphics[width=0.5\textwidth]{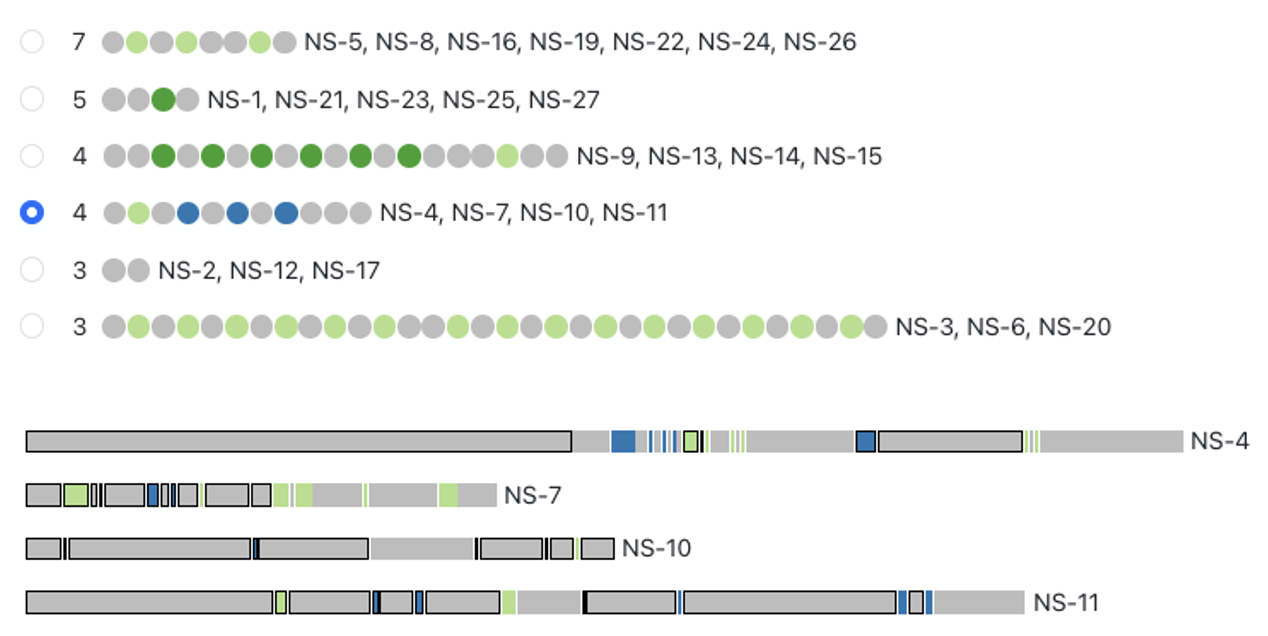}
    \caption{Six patterns returned by Sequence Synopsis~\cite{chen2017sequence} for native speakers during the collaborative writing stage. Each pattern is a sequence of circles, representing a visual overview of sequences belonging to the cluster. The original sequences of the currently selected pattern are displayed below. They are sequences of rectangles, with event duration encoded by length. Events matched to the pattern are outlined in black. 
    % \leo{add some space or a separator between the six patterns and the individual sequences}\tracy{added}
    % \leo{remove the dropdown menus and turn this figure into a 1-column figure, add the explanations on author type and turn in the caption}\tracy{revised}
    }
    \label{fig:cluster-1}
    \Description{Six writing behavior patterns for native speakers during collaborative writing, generated by Sequence Synopsis. Each pattern is shown as a sequence of circles representing a summary of clustered sequences. Below, original sequences are displayed as rectangles, with event durations encoded by rectangle length. Events that match the pattern are outlined in black.}
\end{figure}

We then designed an initial visualization to show the clustering and pattern mining results from these methods. Fig.~\ref{fig:cluster-1} shows the result produced by Sequence Synopsis for NS' behaviors in the collaborative writing stage. There are six clusters, and for each cluster, Sequence Synopsis returns a representative pattern depicted as a sequence of circles. The number before each pattern is the cluster size, and the authors' IDs in the cluster are displayed after each pattern. To show the sequences in the cluster, users can click the radio button before each pattern (currently the 4th pattern is selected). In the raw event sequences displayed below the patterns, each event is a rectangle, where the color denotes the event type, and the length is the duration of the event. We highlight the event rectangles in black outlines if they are reflected in the pattern. We also implemented a similar interface for method II.

\bpstart{Visual summary}
Though the patterns returned by Sequence Synopsis preserve the ordering of events in the sequences, the communication researchers expressed difficulty in interpreting such patterns \add{and wonder whether it's possible to start with something more intuitive, for example, they suggest starting with one author's sequence and building clusters based on similar authors.} \add{Besides,} it was unclear how each pattern differed from one another and how the algorithm performed the clustering. Take the sequences in Figure~\ref{fig:cluster-1} for an example, the second and third cluster both feature a subsequence of \colorbox{WColor}{\small Writing}-\colorbox{PSColor}{\small \small Passive Search}. However, it is unclear how these two clusters differ from each other. The same issue can be found in the first and the last clusters as well, where both clusters exhibit repetitive \colorbox{WColor}{\small Writing}-\colorbox{ASColor}{\small Active-Search} subsequences.  

\bpstart{Cluster membership}
Since we have two sets of clustering results returned by Sequence Synopsis and hierarchical clustering, we let communication researchers switch to different results via a radio button. However, this caused great confusion. Even if we keep the number of clusters $K$ the same for both methods, the clustering results returned by hierarchical clustering and Sequence Synopsis are not the same, and communication researchers are not sure which one to trust more. Besides, there are no explanations for why the sequences are grouped into each cluster or the differences between the clusters. 

To summarize, the communication researchers encountered the following interpretation challenges:

\begin{itemize}
    \item \textbf{C1:} two clustering methods' output differ.
    \item \textbf{C2:} lack of explanations of why the sequences are assigned to a particular cluster.
    \item \textbf{C3:} the extracted patterns are not easily interpretable.
\end{itemize}

\subsection{Strategies to address the interpretation challenges}
To address the above challenges, we devised several strategies, including ensemble and interactive clustering (\textbf{C1}), visualizing sequence-level information for clustering rationales (\textbf{C2}), and using large language models (LLM) to generate text summaries (\textbf{C3}).

\subsubsection{Support ensemble and interactive clustering (C1)}

To address \textbf{C1}, we decided to show the consensus and discrepancies in the results produced by different methods, inspired by the idea behind ensemble clustering~\cite{vega2011survey}. 

We obtained multiple clustering results with varying numbers of clusters, ranging from 2 to $N$, where $N$ represents the maximum number of patterns identified by Sequence Synopsis. We use Sequence Synopsis as a constraint because hierarchical clustering can produce as many clusters as the data points. Then we evaluate the clustering results of both Sequence Synopsis and hierarchical clustering using the same number of clusters. Given two clustering results $A$ and $B$, and a cluster number $K$ ($ 2 <= K <= N$), we try to \add{match each cluster in $A$ to a cluster in $B$ in a way that maximizes the total overlap between cluster members}. We used the Hungarian algorithm~\cite{kuhn1955hungarian} to obtain the \add{assignments with the most overlap.} Then, we only keep assignments when the intersection size is larger than 1 (it is trivial to have a cluster of only one sequence). Therefore, the size of consensus clusters is usually smaller than $K$. For unclustered sequences, we treat them as singletons. 

\begin{figure}
    \includegraphics[width=0.5\textwidth]{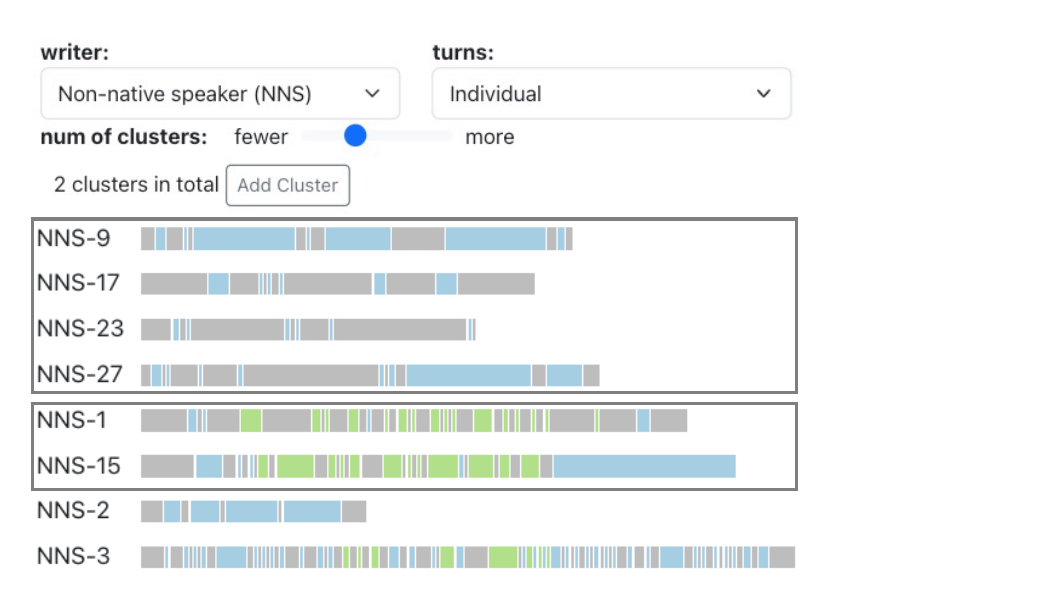}
    \caption{Consensus of clusters: sequences assigned to the same cluster by both Sequence Synopsis and hierarchical clustering are in the box; other sequences are outside of the box. Users can also change the number of clusters by dragging the slider, or manually add new clusters.}
    \label{fig:cluster-2}
    \Description{Sequences assigned to the same cluster by both Sequence Synopsis and hierarchical clustering are in rectangles; other sequences are outside of the rectangle. It also has a cluster slider and an “Add Cluster” button.}
\end{figure}

Figure~\ref{fig:cluster-2} shows a revised version of the visualization, where the sequences enclosed within a rectangle box have cluster assignments confirmed by both SequenceSynopsis and hierarchical clustering methods. In contrast, sequences without a consensus (i.e., NNS-2, NNS-3) are not enclosed, indicating they are singletons. \add{Singletons are expected since the two methods leverage different computational techniques. Singletons remind communication researchers to give additional consideration to these authors, as computational methods differ in their cluster memberships. To help users assign singletons to clusters and revise existing cluster memberships, we provide a slider to adjust the number of clusters and support manually rearranging authors across clusters via drag-and-drop. Furthermore, to assist analysts in finding similar authors based on an author of interest, we also implemented a recommendation feature: we calculated the similarity between two sequences by summing over their sequence-level Levenshtein distance with the Levenshtein distance between their Sequence Synopsis~\cite{chen2017sequence} cluster patterns, and recommended top five to users.} 

\delete{In addition, we also included a slider for users to adjust the number of clusters and support manually rearranging sequences across clusters via drag-and-drop. However, users need more detailed information, such as the similarity between two sequences, to determine whether they belong to the same cluster.}

\subsubsection{Visualize sequence-level information for clustering rationales (C2)}

To help users understand why sequences are clustered (\textbf{C2}) and assess the similarities and differences between sequences within and across clusters, we devised two solutions: one focuses on local information of individual sequences, while the other focuses on the global context:

\begin{itemize}
    \item \textbf{Local information:} for individual sequences, we improve the visualization design to support users in comparing pairs of sequences visually. 
    \item \textbf{Global context:} for all sequences, we reveal their pairwise distances to support users in understanding the overall distribution of sequences in terms of pairwise similarity.
\end{itemize}

\bpstart{Visualizing local information of individual sequences}
As shown in Figure~\ref{fig:cluster-2}, our earlier visualization design of an individual sequence presents the event sequences as it is, where colors encode the event types and the rectangle length encodes the duration. Such a design preserves all the information in the raw data, and communication researchers quickly conclude that NNS usually exhibit much more fragmented workflows than NS, frequently alternating between writing and other events. In contrast, NS usually allocates large chunks of uninterrupted time dedicated to writing. However, it is hard to generate additional insights. Therefore, we considered several design variants for visualizing individual sequences: trees, transition matrices, and arc diagrams. 

\begin{figure}
    \centering
    \includegraphics[width=0.5\textwidth]{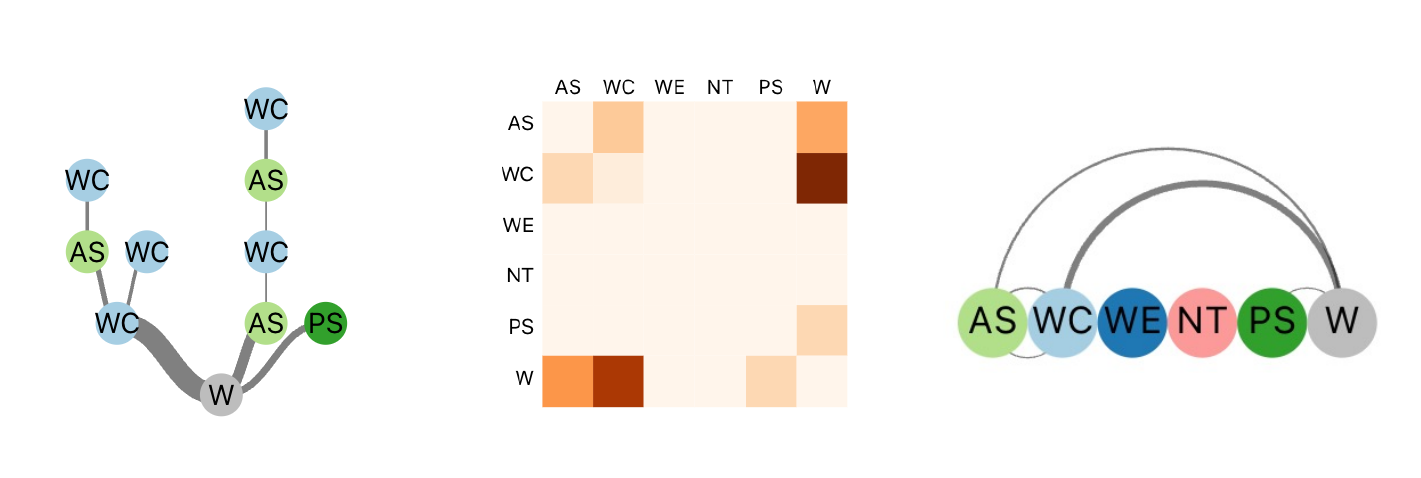}
    \caption{Design variants: tree, transition matrix, and arc diagram (final design) to visualize sequence information for author NNS-15. 
    }
    \label{all-variants}
    \Description{Three design variations for visualizing sequence-level information of an author’s writing behaviors: tree diagram, transition matrix, and arc diagram. The arc diagram is the final selected design.}
\end{figure}

\begin{itemize}
    \item \textit{Variant II: tree.} 
    As communication researchers are interested in comparing NS and NNS' events before \colorbox{WColor}{\small Writing}, we extracted all unique subsequences ending with \colorbox{WColor}{\small Writing}. As depicted in Fig.~\ref{all-variants}, each tree node represents an event, and each edge denotes a transition, with the edge's thickness reflecting the frequency. While communication researchers appreciated the completeness of the tree visualization, they noted some redundancies, e.g., excessively extended tree branches like ``\colorbox{WCColor}{\small WC}-\colorbox{ASColor}{\small AS}-\colorbox{WCColor}{\small WC}-\colorbox{ASColor}{\small AS}''.

    \item \textit{Variant III: transition matrix.} 
    %In this design, 
    To mitigate the issue of redundant sequences, we adopted a transition matrix, where each row/column is an event, and each transition starts from the row and ends at the column. The cell's darkness signifies the normalized transition frequency. As illustrated in Fig.~\ref{all-variants}, it's easy to identify common event pairs, e.g., from \colorbox{WColor}{\small Writing (W)} to \colorbox{ASColor}{\small Active-Search (AS)} and vice versa. However, the matrix is sparse as frequent transitions concentrate on a few cells, resulting in underutilized space. 

    \item \textit{Variant IV: arc diagram.} For the final design, we chose an arc diagram. We borrowed the event node design from the tree visualization, but instead of arranging them in a tree structure, we put all the nodes on the same row and connected them with arcs of different thicknesses, indicating the transition frequencies. The communication researchers like the simplicity and compactness of the design, as it is easy to spot which events precede \colorbox{WColor}{\small Writing}. Besides, unlike the transition matrix, where it is challenging to eliminate empty cells, we can easily conceal unwanted arcs. For example, we removed outgoing edges from \colorbox{WColor}{\small Writing}, as we are interested in events that precede each writing action instead of after. 
    
\end{itemize}

\bpstart{Global pairwise distance of sequences}
After improving the visualization for individual sequences,
we helped users compare sequences globally via pairwise distances. For hierarchical clustering, we computed the Levenshtein distance between sequences and normalized it by the sequence length. Since Sequence Synopsis does not return distance scores, we used the Levenshtein distance between patterns as a proxy. Therefore, if sequences belong to the same cluster according to Sequence Synopsis, their distance is 0. We plot it on a 2D scatterplot, with each sequence represented as a dot, and the sequence of interest is placed at the origin.

\begin{figure*}[h!]
    \centering
    \includegraphics[width=\textwidth]{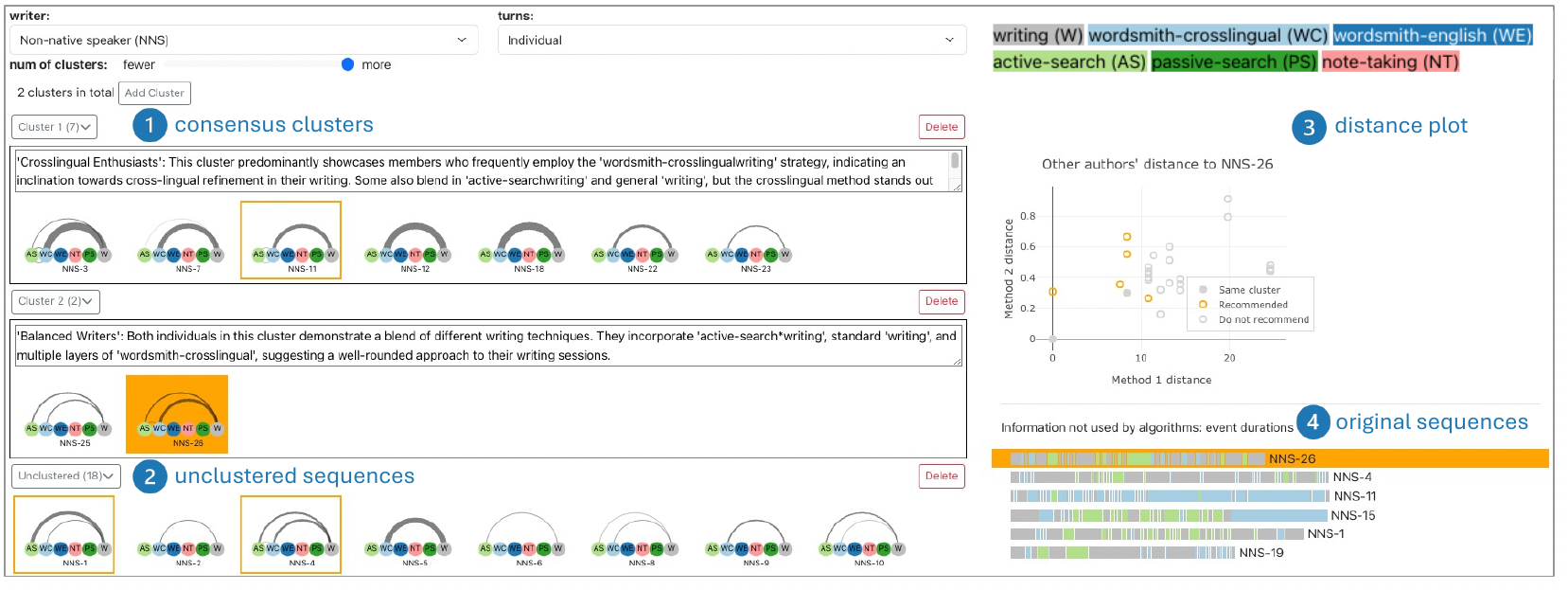}
    \caption{
 The clustering panel of \name{}. \textcircled{1} shows consensus clusters of authors returned by Sequence Synopsis~\cite{chen2017sequence} and hierarchical clustering; \textcircled{2}: unclustered sequences.
  When users select an author, its background turns orange, and recommended authors are highlighted in orange outlines. On the right, a 2D scatterplot (\textcircled{3}) shows the current author's behavior distance between other authors. \textcircled{4} shows detailed sequence information.
  }
  \Description{The clustering panel of COALA. The top left displays consensus clusters of authors from Sequence Synopsis and hierarchical clustering, while the bottom left shows unclustered sequences. When an author is selected, the background turns orange, and recommended authors are outlined in orange. The top right features a 2D scatterplot visualizing behavior distances between authors, and the bottom right presents detailed sequence information.}
    \label{fig:comparison-1}
\end{figure*}

\subsubsection{Use LLMs to generate text summaries (C3)}
To address \textbf{C3} (extracted patterns are not interpretable), we drew inspiration from the analysis process of communication researchers. We observed that they described clusters in natural language; for example, they described the visualization in Figure~\ref{fig:cluster-2} as ``\textit{The first cluster mostly shows writing and wordsmith-crosslingual, so the NNS participants here spent most of their time figuring out how to produce writing in English through Japanese. The second cluster features more active search during the writing session even though participants still performed crosslingual language editing at the beginning and end of their sessions. These NNS participants followed a different writing path - they segmented the content (active information search in the middle) and the editing (wordsmith at the beginning and end) aspects of the writing and focused on one task at a time.}''

Recently, large language models (LLM) have shown great promise in data analysis~\cite{brown2020language, ouyang2022training, ma2023insightpilot} and LLM is found to have a high degree of agreement with human coders in thematic analysis~\cite{dai2023llm}. Therefore, we leveraged LLM to generate text descriptions. We first tried capturing a screenshot of each cluster of arc diagrams (Figure~\ref{all-variants}) and prompted GPT-4V~\cite{VisionOp27:online} with an explanation of the color encodings to describe each cluster. For example, we used the following prompt: ``\textit{The figure contains several event sequences, each showing an author's writing-related behaviors. There are six types of events: Active-Search, Wordsmith-Crosslingual, Wordsmith-English, Note-Taking, Passive-Search, and Writing. Each colored node is an event type, and you can find the event type in the colored legend. The arc thickness is the transition frequency of two events. Please name this cluster and provide a brief description.}'' 

However, GPT-4V sometimes ignored faint arcs between two nodes. To ensure GPT captures all transitions, including rare ones, we provided the transition data in JSON format, which was used to generate the arc diagram. Each entry contains the source event, the destination event, and the normalized frequencies. For example, source: \colorbox{WCColor}{\small Wordsmith-Crosslingual}, destination: \colorbox{ASColor}{\small Active-Search}, frequency: 0.25. Communication researchers found explanations returned by GPT fascinating and intuitive, and adopted some descriptions in their analysis. For example, GPT-4V calls one cluster ``versatile writers'' and explains that the cluster balances events between \colorbox{WCColor}{\small Wordsmith-Crosslingual} and \colorbox{ASColor}{\small Active-Search}, which suggests writers are comfortable with both actions without a dominant one. 

\add{Our prompting strategy to generate clusters could be generalized for other sequence datasets, and could be a plug-in for many existing visual analytics systems. We also found in later user study sessions 
%\zinat{sessions?} 
that users borrow words from LLM-generated summaries during the analysis, especially for users new to the dataset.}

%%%% 5-methods.tex ends here %%%%

%%%% 6-interface.tex starts here %%%%

\section{\name{}}

\begin{figure*}
    \centering
    \includegraphics[width=\textwidth]{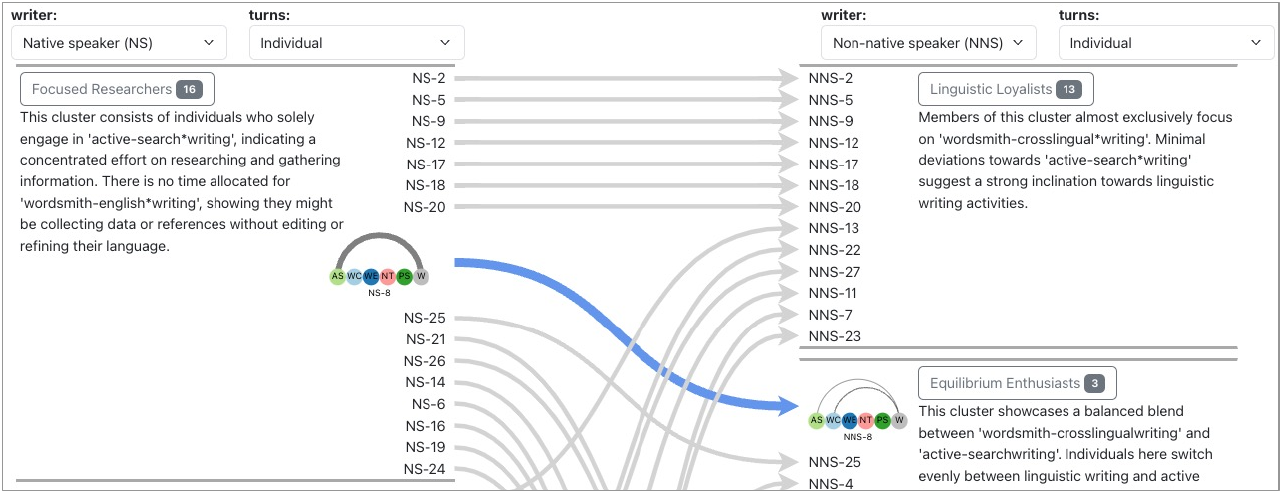}
    \caption{The comparison panel of \name{}. It shows a two-panel layout: each has its dropdown menu for users to configure the author and writing stages. An arrow connects authors from the same team on both panels. Users can click an arrow to directly compare the two sides' arc diagrams.}
    \Description{The comparison panel of COALA, featuring a two-panel layout where users can configure authors and writing stages via dropdown menus. Arrows connect team members across both panels, enabling users to click an arrow to directly compare their arc diagrams.}
    \label{fig:comparison-2}
\end{figure*}

Integrating all these strategies, we built COALA, a visual analytics tool to \textbf{co}mp\textbf{a}re col\textbf{la}borative writing behaviors of native and non-native English authors. It has two tabs: one for inspecting and modifying the clustering results (Figure~\ref{fig:comparison-1}) and the other for comparing clustering results for authors of interest (Figure~\ref{fig:comparison-2}).

The first tab (Figure~\ref{fig:comparison-1}) supports refining sequence clusters and summaries. After selecting authors and writing stages in the dropdown menu, it displays consensus clusters of sequences by Sequence Synopsis and hierarchical clustering (\textcircled{1}) and unclustered sequences (\textcircled{2}). Users can drag the slider to change the number of clusters, add or delete clusters, revise cluster descriptions, and drag and drop arc diagrams directly across clusters. To facilitate the refinement process, when users select an author's arc diagram, its background turns orange, and recommended similar authors outside its cluster are highlighted in orange outlines. Currently, an author in the second cluster (orange background) is selected, and several authors outside the cluster are recommended. On the right, a 2D scatterplot (\textcircled{3}) shows the selected author's sequence distance between other authors according to Sequence Synopsis (method 1) and hierarchical clustering (method 2), and users can hover over each dot to see the sequence ID and distance. Users can also hover over \textcircled{4} to inspect the event types and durations.

The second tab (Figure~\ref{fig:comparison-2}) supports directly comparing collections of sequences and individual sequences. It shows a two-panel layout; each panel has its dropdown menu for users to select collections of sequences. Currently, we select NS-individual on the left panel and NNS-individual on the right panel. An arrow connects authors of the same team. Authors are organized by clusters in Figure~\ref{fig:comparison-1} and sorted by minimizing edge crossings. Once the user clicks an arrow, arc diagrams will appear on both panels. For example, now it shows that NS-8 transitions frequently between \colorbox{ASColor}{\small Active-Search} and \colorbox{WColor}{\small Writing}, while NNS-8 has a balanced \colorbox{ASColor}{\small Active-Search} and \colorbox{WCColor}{\small Wordsmith-Crosslingual} activities before \colorbox{WColor}{\small Writing}, highlighting NNS' effort in information gathering and also translation. \add{For cluster-level comparisons, since the arrows are sorted, users can observe the flow directions, similar to a Sankey diagram.}

%%%% 6-interface.tex ends here %%%%

%%%% 7-study.tex starts here %%%%

\section{Validation: User Studies}

The validity of our design is rooted in the user-centered design process reported in the previous sections. We also organized two user studies to evaluate \name{}. 
%\leo{Summarize the two user studies and their goals. That is: why two user studies?}
\add{The first, a focus group session with two existing and two new communication researchers, examined the usage of \name{} in real-world setting; the second, individual sessions with 8 graduate students, evaluates the effectiveness of \name{} for general users who work more independently.}

\subsection{User study setup}

\subsubsection{Focus group (\add{N=2+2})} 
We organized a focus group session with \add{the two existing} communication researchers \add{along with two of their colleagues}\delete{(N=4)}. \add{All four participants belong to the same research group. The two new researchers} are familiar with the multilingual communication research but are unfamiliar with the dataset details, nor have they seen \name{} before. Though our collaborators are informed about the methods and individual visualizations, they also have not used \name{}. Therefore, we conducted a 1-hour study session with all communication researchers (N=\add{2+2}). \add{The goal of the group study is to use COALA in a realistic work setting to make sense of a dataset.} 
%\leo{Change to: The goal of the group study is to use COALA in a collaborative setting to make sense of a dataset.} 
In the beginning, we introduced the dataset and analytic tasks (\textbf{T1} and \textbf{T2}), and then we gave a tutorial on how to use the tool. We deployed \name{} online, and each researcher accessed the tool on their laptop. 
%\leo{So the researchers are using the tool in parallel?}\tracy{yes} 
They were encouraged to think aloud and discuss their findings during the session. We took notes during the session, and after the group session, we also invited them to write down their findings in a shared document.

\subsubsection{\add{Individual user studies (N=8)}}
\add{Besides the focus group, we also expanded the evaluation scope by recruiting eight graduate students with related research backgrounds (communication: 2, second-language education: 2, CSCW: 2, political science: 1, natural language processing: 1). All participants have first-hand collaborative writing experiences and are interested in understanding collaborative writing behaviors. Among them, four are NS, and four are NNS. To better evaluate the usability and design effectiveness of \name{}, we set up individual user study sessions with them.} 
%\tracy{To better evaluate the usability and design effectiveness of \name{}}, we conducted individual sessions with each participant. \leo{We should have better reasons for doing individual sessions than scheduling conflicts. I remember Yimin proposed some arguments for justifying both individual and group studies. We need to state the goals for each type of study clearly.}\tracy{added. For the group study, I emphasized that we replicate communication researchers' workflow; for the individual study: better evaluate the usability and design effectiveness.}

\add{The procedure is similar: first, participants read a background introduction of the dataset (we prepared a shortened and simplified version of section ~\ref{domain-background} and section~\ref{data-abstract}). We then demonstrated how to use \name{} to cluster similar authors and compare their writing behaviors. Participants were asked to complete two tasks: analyzing authors' behaviors across two dimensions—writing stages (individual vs. collaborative) and author types (NS vs. NNS). Participants can either write their findings in \name{} or describe them verbally. Participants are given one hour to complete the tasks. Participants were asked to think aloud during the study, and after they completed the tasks, we conducted a brief interview to ask them questions about the effectiveness of \name{} and the quality of model-generated results. We chose not to collect the ratings quantitatively due to the limited sample size and the lack of a baseline tool for comparison. Instead, we focus on gathering more qualitative feedback. All except one participant completed the tasks on time, and every participant was paid with a \$20 Amazon gift card. The study is IRB-approved.}

\subsection{Effectiveness of COALA}
\subsubsection{Focus group (N=2+2). \name{} facilitate discussions}
\add{All \add{expert users}\delete{participants} had no difficulty in using \name{}, and they were surprised to see that they focused on different aspects of the dataset during the analysis. For instance, one focused on the durations in the event sequence, and the other focused on the event transitions. This divergence triggered experts' discussion of why one aspect of collaborative writing was important. Ultimately, they concluded that \name{} offered multiple angles to analyze the dataset they would otherwise miss due to their intuitions.}

\subsubsection{Individual user studies (N=8). \name{} helps uncover insights}
\add{Among the seven participants who completed the study, three found the arc diagram design more effective, two preferred the sequence diagram, and two considered both effective. Participants prefer the arc diagram mostly because it summarizes the event sequences. For example, P8 appreciated how the arc diagram \textit{``compresses information''}. In contrast, participants preferred the sequence diagram because it encoded the duration information (P5).} 
%\leo{If they think something is not effective, what is the reason?}\tracy{mostly because it does not have the information they want, e.g., time duration} \leo{add these in}

\add{Regarding the model quality, five out of seven participants considered the initial clusters obtained by the consensus methods to be high quality and encouraged unbiased exploration. For example, P7 said, \textit{``It's a really good starting point, because if you were to analyze this by yourself without any ideas...you can fall into typical stereotypes that...an AI model would not.''} Furthermore, six participants found the recommendation feature helpful, using it as a starting point for the analysis (P8), a voting mechanism (P6), and a verification method (P7)}.

\add{Participants also made suggestions to improve model-generated results, including more details about how the recommendation methods work (P4), more fine-grained clusters to start with (P5), and a 3rd recommendation method based on time duration (P6)}. 

%\leo{This subsection should be very important, but I feel it is a bit too brief.}\tracy{I may be able to extract more users' quotes, but the main conclusions may not change}
%\zinat{Should we also include some commentaries (if any) the participants had on the entire system? For example, if any of them mentioned that this kind of system will be helpful to understand writing behaviours more efficiently than simple data analysis or so on?}\tracy{I did not ask such questions. Added a summary.}

\add{Notably, none of the participants had analyzed multilingual collaborative writing datasets before. Despite their unfamiliarity with the dataset and \name{}, most of them were able to familiarize themselves with \name{} and complete the analysis tasks. By leveraging the model-generated results and data visualization designs, they uncovered insights similar to those of expert users and connected the findings with their own experience. This outcome highlights the general usability and design effectiveness of \name{}.}

%\leo{Experts (N=2)}

%\leo{General Users (N=2+8)}

%\subsection{Inter-Participant Dynamics from Focus Group}

\subsection{Findings: collaborative writing patterns}\label{group-findings}
%\leo{add some introductory text to explain these are the findings from the participants by using COALA.}
\add{Here we aggregated participants' findings by using \name{}, including the dataset itself and their reflections on their own collaborative writing experience.}
\subsubsection{Individual stage patterns}\label{sec:1}
NS and NNS are asked to write independently before collaborating in this experiment to avoid free-riders. Communication researchers observed distinct behaviors during the individual stage.

\bpstart{NS research extensively} 
% \leo{It's good to start each paragraph with a summary of the findings, but use a complete sentence. } \tracy{added} 
Using our tool, communication researchers easily identified several major clusters of NS. One cluster is characterized by dominant \colorbox{ASColor}{\small Active-Search}-\colorbox{WColor}{\small Writing} behaviors, indicating they actively incorporated external information into their writing. Another cluster has more activities, besides \colorbox{ASColor}{\small AS}-\colorbox{WColor}{\small W}, they also engaged in extensive paraphrasing activities: \colorbox{WEColor}{\small Wordsmith-English}-\colorbox{WColor}{\small Writing}.
% , implying paraphrasing activities. 
Besides these two major clusters, there are also uncommon behaviors, e.g., one NS only displayed \colorbox{WEColor}{\small WE}-\colorbox{WColor}{\small W} behavior without relying on external information; one
NS went even further, solely engaged in writing, degenerating the arc diagram into a stick on the \colorbox{WColor}{\small W} node.

\bpstart{NNS encounter costly bilingual content production} Similarly, communication researchers identified several clusters of NNS. In one cluster, \colorbox{WCColor}{\small Wordsmith-Crosslingual}-\colorbox{WColor}{\small Writing} dwarfed any other behaviors, implying NNS had significant usage of their native language to produce writing in English. Another cluster has more \colorbox{ASColor}{\small Active-Search}-\colorbox{WColor}{\small Writing} behaviors, showing NNS also incorporated external information into their writing. Different from NS, several NNS also displayed \colorbox{ASColor}{\small AS}-\colorbox{WCColor}{\small WC}, meaning NNS also used their native language to transfer content from their content-related search to their writing output. 
% The thickness of the arc makes it easy to compare patterns. 
Among the NNS who have engaged in both \colorbox{WCColor}{\small WC}-\colorbox{WColor}{\small W} and \colorbox{ASColor}{\small AS}-\colorbox{WColor}{\small W}, half are dominated by \colorbox{WCColor}{\small WC}, and the rest are dominated by \colorbox{ASColor}{\small AS} or are balanced. Notably, none of the NNS engaged with \colorbox{WEColor}{\small Wordsmith-English}-\colorbox{WColor}{\small Writing} during the individual writing stage, highlighting the bilingual nature of NNS' writing process, as their limited English proficiency may have dissuaded them from writing directly in English. Compared to NS, many more NNS (n=10) only engaged in \colorbox{WCColor}{\small WC}-\colorbox{WColor}{\small W} without relying on external information. This pattern hinted that NNS have dedicated much of their time to the costly process of writing in English - generating content in their native language first and then translating the content to English, either by themselves or by leveraging support from cross-lingual dictionaries and translation tools.

\subsubsection{Collaborative stage patterns}\label{sec:2}

Communication researchers 
% \leo{what does "it" refer to?}\tracy{added} 
found more diverse actions during collaborative stages, driven by co-authors' need to exchange information, refine text, and communicate with each other. For example, communication researchers observed multiple NS (n=8) and NNS (n=10) engaged in \colorbox{PSColor}{\small \small Passive-Search}, indicating active information sharing. NS and NNS also display different distributions of behaviors in the collaborative stage.

\bpstart{NS shoulder more editing responsibilities} Transitions between \colorbox{ASColor}{\small AS}-\colorbox{WColor}{\small W} and \colorbox{WEColor}{\small WE}-\colorbox{WColor}{\small W} still have a strong presence, though the balance has shifted, with \colorbox{WEColor}{\small WE}-\colorbox{WColor}{\small W} increases, and 
\colorbox{ASColor}{\small AS}-\colorbox{WColor}{\small W} decreases. The shift towards editing indicates that NS is responsible for editing both parties' English expressions.

\bpstart{NNS gain more bandwidth for other tasks} Communication researchers witnessed an uptick in both \colorbox{WCColor}{\small WC}-\colorbox{WColor}{\small W} and \colorbox{ASColor}{\small AS}-\colorbox{WColor}{\small W}, suggesting NNS actively interpreting and incorporating NS-contributed content. In addition, more than a third of NNS engaged in \colorbox{PSColor}{\small \small PS}-\colorbox{WColor}{\small W}. These observations suggested that when co-writers completed initial drafts and transitioned to a collaborative editing stage, NNS were partially relieved from the demanding task of producing English content. Thus, they had more bandwidth to perform other tasks, such as \colorbox{ASColor}{\small AS} to enrich the joint writing further. 

\subsubsection{Echo with first-hand experience}
\add{In the individual user study sessions, besides the findings similar to the ones in the group study session, our participants also compared their findings through the lens of their collaborative writing experience. For example, P6, a native English speaker, noted that too many NS \textit{``just spitball on the fly without actually doing any research''} in this dataset, which aligned with his previous collaborative writing experience.  In contrast, P2 (a non-native English speaker) observed that \textit{``...native speakers become more cautious with their writing...when they are working with people, which is something that I wouldn't expect...if I am working with a non-native speaker of my language. I feel like I know more, so I would feel less need to actually check my writing.''} P4 resonated with the \colorbox{ASColor}{\small Active-Search}-\colorbox{WCColor}{\small Wordsmith-Crosslingual} transition in the arc diagram: \textit{``I was seeing myself do the same thing. I also like search for specific words in Korean, and then translated into English.''} }

\add{Some participants also voiced their suggestions for best practices in collaborative writing. For example, P3, a native English speaker, pointed out that so much time spent on \colorbox{WCColor}{\small Wordsmith-Crosslingual} is \textit{``a waste of manpower, it makes more sense to focus on the ideation and don't care if it comes out ugly if the other person (NS)  can fix it without having to do a lot of wordsmithing.''}}

\subsection{Findings: participants' analysis strategies}
%\leo{Say something about how the strategies are obtained, what data did we collect and how did we turn the data into strategies?}
\add{We also studied participants' analysis strategies. For the group study, we analyzed the notes taken during the study and the document on which participants wrote findings. For the individual study, we analyzed the video recordings, including participants' screens and meeting transcripts.}

\subsubsection{Experts (N=2+2)}
%\leo{More should said on the interaction between the researchers and how COALA promotes the discussion. How do they collaboratively use it? How do they resolve differences in opinion, e.g., one wants higher-level clusters and the other wants lower-level clusters? How does the fact that multiple people use the tool at the same time allow them to examine the data from different perspectives focusing on different dimensions?}
\name{} is pre-populated with clusters returned by methods described in section~\ref{method-section}. Initially, all communication researchers used those clusters 
%(details in section~\ref{method-section}) %\leo{section number}\tracy{added} 
to gain an overview of the dataset. They read the descriptions generated by GPT-4 and gradually started to refine clusters based on their understanding. 

For example, one researcher (E1) merged clusters based on the role of external resources in the writing processes:
% whether the author needs to borrow ideation from external resources, 
authors who use \colorbox{WEColor}{\small \textbf{Wordsmith-English}} or \colorbox{WCColor}{\small \textbf{Wordsmith-Crosslingual}} are classified as \textit{``grammar-based''}, as they only need help with the expression, but not ideation;  authors who leverage \colorbox{ASColor}{\small Active-Search} and \colorbox{PSColor}{\small \small Passive-Search} are classified as \textit{``research-heavy''}, as such authors are still in the process of finding ideation. 

On the opposite, another researcher (E2) broke down clusters into sub-clusters by examining the duration and transition frequencies of raw sequences carefully (Figure~\ref{fig:comparison-1}.\textcircled{4}); for example, NS-20 and NS-22 were initially clustered into the same cluster by our algorithms as their behaviors are dominated by \colorbox{ASColor}{\small Active-Search}, however, after examining the raw sequences, E2 found that NS-20 usually spent a long time on \colorbox{ASColor}{\small Active-Search} before transitioning to \colorbox{WColor}{\small Writing}, different from NS-22, who rapidly transitioned between those two. Therefore, E2 created a new cluster called \textit{``in the flow''}, and placed authors with less fragmented workflow into this cluster. 

During the refinement, they also use the recommendations (orange border) to guide them in finding similar authors. After refinement, they moved to the comparison panel (Figure~\ref{fig:comparison-2}) and examined individual authors' behavior changes. 
% \leo{These descriptions on their strategies are very low-level, it is unclear what their rationales were when doing these.}\tracy{they're given the same tasks: T1 and T2}

\delete{All participants have no difficulty in using \name{}. However, they identified a limitation with adjusting the number of clusters using only the slider: as the number of clusters increases, some clusters' descriptions are highly similar. They suggested that \name{} could automatically merge similar ones and retain only those with distinct text descriptions.}

\subsubsection{General users (N=8)}
\add{Similar to focus group participants, participants in individual sessions also started by understanding the pre-populated clusters and text descriptions. }

\bpstart{Clustering strategies} 
% \leo{Start by summarizing if we find the same clustering strategies as the group study, or if we find something different.}
\add{We found participants in individual sessions follow similar high-level clustering strategies of E1 and E2.} \add{Similar to E1, four participants also clustered authors by grouping multiple activities. For instance, P3 and P5 assigned authors performing more than two activities to a \textit{``multi-tasking''} cluster; P3 assigned authors with \colorbox{ASColor}{\small AS} or \colorbox{PSColor}{\small \small PS} 
% \leo{does this mean authors with AS and authors with PS, or authors with both?}\tracy{use or instead} 
to the same cluster, as such behaviors implied that the authors 
% \leo{``they'' refer to the authors or the events?}\tracy{clarified} 
incorporated external information.}

\add{Similar to E2, three participants refined the clusters with more fine-grained criteria. For example, P4 created two clusters for authors exhibiting dominant \colorbox{ASColor}{\small AS}-\colorbox{WColor}{\small W} behaviors and labeled them as \textit{``comfortable/uncomfortable with searching information in English''} based on the absence of \colorbox{ASColor}{\small AS}-\colorbox{WCColor}{\small WC}. P7 and P8 broke down clusters by considering the events' time durations, for example, P7 differentiated authors with \colorbox{WCColor}{\small WC}-\colorbox{WColor}{\small W} behaviors by the event durations and labeled authors spent significant time on \colorbox{WCColor}{\small WC} as \textit{``long-term crosslingual enthusiasts''}. Similarly, P8 created several clusters with names like \textit{``high/low active-search''} and \textit{``strong/weak passive-search''} based on the durations.}

% \leo{are there people who employ both strategies? they don't seem to be mutually exclusive.}\tracy{I don't remember having seen such participants}

\add{Participants also interpreted the same behavior differently. For instance, for authors who wrote extensively without 
engaging in other activities, participants described them as \textit{``have lots of knowledge and can pull citations from their memory''} (P4), \textit{``stick to minimal approaches''} (P5), \textit{``YOLO''} (P6) or \textit{``confident''} (P7).}

\bpstart{Reliance on the recommendation feature}
\add{Compared to E1 and E2, participants in the individual sessions rely more on the recommendation feature to discover similar authors. Some participants used it as a starting point to narrow the candidate pool; others used it to verify their assumptions. For example, P7 identified similar authors visually and then clicked the author of interest to see whether the recommendation confirmed her assumption. P6 employed the recommendation feature as a voting mechanism: when adding a new author to an existing group, P6 cross-checked the recommendation of multiple existing authors and only picked those endorsed by multiple existing cluster members.}

\add{Similarly, after refining clusters, participants moved to the comparison panel to get an overview of the authors' behavior changes and clicked authors of interest for more fine-grained comparison.}

%%%% 7-study.tex ends here %%%%

%\input{7-group-study}
%\input{8-individual-study}
%\input{8-validation-coco}
%\input{9-lessons-learned}
%\input{10-conclusion}

%%%% 8-discussion.tex starts here %%%%

\section{Discussion}

In this section, we reflect upon the lessons learned from understanding collaborative writing processes through visual analytics and the implications for AI-assisted collaborative writing tools.

\subsection{Reflections on developing visual analytics tools}
% \bpstart{Intuitive details vs. confusing overviews} 
\bpstart{Information-seeking and visual analytics mantras have limitations in guiding tool design}
When we started working with the communication researchers, we aimed to create an overview of authors' behaviors, following the visual information-seeking mantra ``overview first, zoom and filter, then details-on-demand''~\cite{shneiderman2003eyes}. Therefore, we chose Sequence Synopsis~\cite{chen2017sequence}, which clusters sequences and generates an overview pattern for each cluster. However, as discussed in section~\ref{interpretation-challenge}, communication researchers found such overviews confusing. Instead, they asked whether it was possible to start with one author's sequence and manually build clusters based on similar authors. This request directly contradicted the mantra, and we realized that the introduction of computational models resulted in interpretation challenges, and an overview would only be meaningful if users could reliably interpret it. Since it is tedious to inspect and cluster individual sequences manually, we kept the automatic clustering results but addressed the interpretation challenges with ensemble and interactive clustering and large language model (LLM)-generated text summaries. 
% During the development, we realized the deviation from the classic information-seeking mantra is due to the introduction of computational models, which may generate results that are not interpretable for domain experts. 
With the advances of AI, we expect more complex computational models to be incorporated into analysis. Although the visual analytics mantra (``Analyze first, show the important, zoom/filter, analyze further, details on demand'') proposed by Keim et al. \cite{keim2008visual} tries to address the importance of analysis, what constitutes ``the important'' is highly dependent on the computational models and the tasks involved. Therefore, it is critical to emphasize the interpretability of model results and build a shared representation between analysts and models~\cite{heer2019agency}.

\bpstart{Visualizing ensemble clustering methods requires consolidating results} Clustering simplifies the analysis by grouping similar data points, but the variety of clustering methods and clustering results raised doubts among our collaborators about the output reliability. At first, we simply used radio buttons to toggle between methods, inadvertently presenting each method as interchangeable black boxes. However, having an agency over black boxes does not gain users' trust. Therefore, we visualized the uncertainty explicitly using a bounding box: sequences assigned to the same cluster by different methods were clustered, while discrepancies resulted in singletons. Besides, we also visualized the distance between different authors in a scatterplot to allow communication researchers to explore the similarity space intuitively.
% \leo{what is the lesson here that can be applied to other situations?}\tracy{added} 
Our lessons show that mere juxtaposition is not enough when visualizing multiple models' results; instead, we should support analysts in interpreting multiple models' results without model-specific knowledge.

\bpstart{Caution needs to be exercised when updating model results based on user feedback} Our communication researchers think the automatically generated clusters are a useful starting point, yet still require revisions. Therefore, we implemented several user interactions, such as adjusting the number of clusters and rearranging cluster members. To minimize manual operations, we further suggested updating the recommendation interactively, e.g., if an author was moved from one cluster to another, then both clusters' descriptions would change, and the distance function would be revised. However, our collaborators found this distracting, preferring a more stable interface and manual revision of AI-generated results. Though incorporating users' feedback can improve the quality of model results~\cite{cohn2003semi}, it also requires more thoughtful inputs from users and imposes extra cognitive load to examine the updated interface~\cite{smith2018closing, strobelt2021genni}. Therefore, future tools should carefully balance the benefits of refined models and the additional burden on users.

\subsection{Design implications for AI-assisted collaborative writing tools}

With the recent advances of large language models (LLM), researchers have proposed multiple AI-assisted writing tools~\cite{lee2024design, dang2022beyond, zhang2023visar, laban2023beyond, dang2023choice, reza2024abscribe, sun2024reviewflow, kim2024diarymate, lee2022coauthor}, some are tailored for non-native speakers~\cite{ito2023use, kim2023towards, chen2024worddecipher}. However, we have not yet seen AI-assisted writing tools targeting collaborative writing, potentially due to the lack of understanding of the dynamics of collaborative writing. Here we propose several potential features derived from our study.

\bpstart{Detect diverse contribution types}
Traditional collaborative writing tools~\cite{viegas2004studying, wang2015docuviz} for tracking authors' contributions in a collaborative document usually rely on word count. In our initial exploration (Appendix), we also found that NS contributed more text, and our collaborators mentioned that sometimes NS complained that NNS wrote too little. However, according to the communication researchers, there are diverse contribution types that are not necessarily manifested in word counts. Tools that quantify multi-dimensional contributions will help teammates understand each other's contributions and foster team dynamics. For example, NNS could contribute to the overall writing by providing an idea, which was expanded and refined by NS. Other contributions, such as scaffolding and improving the document flow by reorganizing, are also equally valuable. Automatically detecting such contributions requires tracking and deeply understanding the document, which may be potentially feasible by leveraging LLM.

\bpstart{Reduce context switching for NNS}
NS and NNS spent about the same time on this lab study. However, the way they spent their time is notably different. NNS spent much more time transitioning between \colorbox{WCColor}{\small Wordsmith-Crosslingual} and \colorbox{WColor}{\small Writing}, resulting in a more fragmented workflow. In contrast, NS usually dedicated extended time to writing, engaging in focused and uninterrupted work, as highlighted by Newport's concept of ``deep work''~\cite{newport2016deep}. To improve efficiency for NNS, we suggest that future collaborative writing tools introduce features to reduce multilingual context switching. For example, code editors like VSCode~\cite{vscode} support generating code by providing instructions in natural language, reducing the time for developers to search syntax of programming languages in external resources. Similarly, an AI-assisted editor could support NNS by providing high-level instructions in their native language and generating text in English.

\bpstart{Provide guardrails for machine translation}
NNS is frequently involved in \colorbox{WCColor}{\small Wordsmith-Crosslingual} during both individual and collaborative writing. However, machine translation is prone to information loss~\cite{vanmassenhove2019lost, mohammad2016translation}, and NNS are usually unaware of it. For example, during an interview conducted by our collaborators, NS-1 mentioned that NNS-1 left a sentence, ``\textit{Why don’t you stop comparing yourself to others and focus on yourself}?'' Though NS-1 perceived an accusatory tone and wanted to change it, NS-1 eventually decided to respect NNS-1's writing and did not revise it. However, it was later revealed that NNS-1 originally wrote it in Japanese and the soft tone was lost in translation. Therefore, future collaborative writing tools could have built-in translators and monitor the tone before and after translation.

\bpstart{Summarize collaborators' activities}
During the collaborative stage, we observed that authors spent some time catching up on the latest changes in the document, for example, reading links provided by co-authors (\colorbox{PSColor}{\small \small Passive-Search}). Though online collaborative writing tools like Google Docs have version history, they only reflect word-level change and do not summarize the semantic meanings of version changes. Therefore, future AI-assisted collaborative writing could support automatically summarizing co-authors' activities and facilitate the syncing up process.

\subsection{\add{Generalizability for understanding diverse collaborative processes}}
\add{Though \name{} is initially designed for a specific dataset that our collaborators have collected, \name{} also has the potential to help analyze more generalized and diverse collaborative processes.}

\bpstart{\add{Diverse collaborative writing settings}} 
\add{Since our collaborators' study is one of the first studies on multilingual collaborative writing, they opted for a simplified setup: one NS and one NNS with clear turn-taking. In the future, communication researchers may conduct user studies with more complex scenarios, e.g., multiple NS and NNS, or NNS from different countries. Since \name{} follows a generalized ``cluster-summarize-compare'' workflow, \name{} can adapt to and process new sequences provided by researchers. Furthermore, the computational methods underpinning \name{} are agnostic to event or author types, enabling researchers to introduce events and author types beyond those in Table~\ref{tab:videoActivityFreq}. For example, if communication researchers want to study how people of varying Wikipedia editing experiences co-write an article, they could leverage a set of events recorded by Wikipedia, such as ``create'', ``add'', ``delete'', ``revert''. In this case, the factor of interest shifts from language proficiency to the authors' Wikipedia editing experience.}

\bpstart{\add{Collaborative processes beyond writing}}
\add{While \name{} focuses on analyzing multilingual collaborative writing, the design principles and computational techniques could extend to a wide range of collaborative processes that generate rich event sequences, such as visual information analysis, knowledge synthesis, and problem-solving.} \add{For instance, Isenberg et al.~\cite{isenberg2008exploratory} examined how different teams behave in an information analysis task and derived a framework for such activities by analyzing temporal sequences. In another study, Isenberg et al.~\cite{isenberg2011co} recruited 15 pairs of participants to solve a problem by exploring 240 digital documents and identified eight collaboration styles. Similarly, Robinson et al.~\cite{robinson2008design} recruited five pairs of geographers and disease biologists to complete a knowledge synthesis task, analyzing the participants' action frequency, time durations, and strategies. Recently, Yang et al. ~\cite{yang2022towards} studied how teams collaboratively engage and perform sensemaking tasks in an immersive environment. These studies highlight the importance of understanding the temporal sequences of participants' behaviors and identifying common patterns—an area where \name{}'s clustering and summarization features could provide significant benefits. Unlike previous methods which rely primarily on manual coding or frequency-based analysis, \name{} offers a data-driven approach to interpret collaborative dynamics, and thus enhances both scalability and granularity in data analysis.}

\subsection{Limitations}
One limitation of \name{} is that it does not visualize text changes during collaborative writing. In the individual user studies, the only participant who struggled to complete the analysis tasks noted that having text alongside event sequences would make the analysis more concrete. While we explored adapting existing text visualizations to our dataset (see Appendix), our focus is on novel approaches to support behavioral analysis. Future tools could seamlessly integrate both aspects to enhance analytical depth.
Another limitation is that \name{}'s comparison feature is designed for dichotomous analysis, such as collaborative vs. individual or NS vs. NNS. For teams without clear binary distinctions, \name{} only supports clustering but lacks dedicated comparison capabilities.

%%%% 8-discussion.tex ends here %%%%

%%%% 9-conclusion.tex starts here %%%%

\section{Conclusion}
To compare native and non-native English speakers' behaviors in collaborative writing, we partnered with two communication researchers to design visual analytics techniques. To address the limitations of existing text and event sequence visualizations, we implemented \delete{a new tool}\add{\name{}} to help communication researchers uncover insights during the collaborative writing process, mainly focusing on addressing interpretation and trust challenges. We validate \delete{tool's}\add{\name{}'s} effectiveness by conducting a focus group study (N=2+2) and \delete{comparing it with another established cohort comparison tool}\add{an individual user study (N=8).} Finally, we share design lessons learned during the development and potential features for AI-assisted collaborative writing tools. We believe this work will significantly contribute to collaborative writing communities, fostering a more diverse and inclusive work environment, especially for non-native speakers. \add{Additionally, researchers aiming to understand other collaborative processes, such as collaborative knowledge synthesis and sensemaking, could also find inspiration in the designs of \name{}.}

%%%% 9-conclusion.tex ends here %%%%

\begin{acks}
We would like to thank user study participants and the anonymous reviewers for the constructive feedback. This work is supported by NSF awards IIS-2239130 and IIS-1947929.
\end{acks}

\bibliographystyle{ACM-Reference-Format}
\bibliography{template}

\appendix
\clearpage
\section{Understanding document modification behaviors}
In this section, we focus on understanding document modification behaviors $E_{on}$. Such behaviors could be inferred by comparing consecutive document versions via the Myers difference algorithm~\cite{myers1986nd}. 

\subsection{Task Analysis}
In our interviews with communication researchers, there are two tasks about analyzing document modification behaviors:

\bpstart{T1: identify editing dynamics} Communication researchers hypothesize that NS and NNS contribute differently to the joint document and would like to know the difference reflected by the evolution of documents.

\bpstart{T2: analyze specific content} Communication researchers are interested in finding frequently co-edited content and seeing how NS and NNS edit them during collaborative writing.

\begin{figure*}
    \centering
    \includegraphics[width=0.7\textwidth]{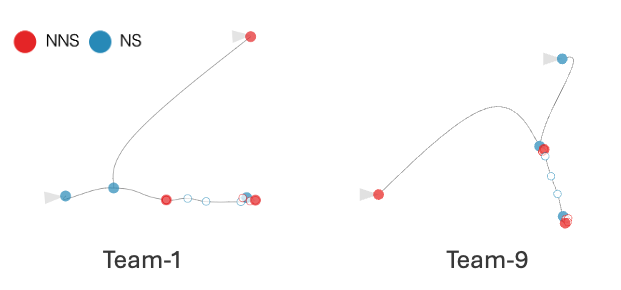}
    \caption{Branched time curves of team-1 and team-9. The gray arrows indicate the first versions written by NS and NNS. Big and filled dots are major versions, and small and hollow dots are intermediate versions.}
    \label{fig:branched-timecurve}
\end{figure*}

\begin{figure*}
    \centering
    \includegraphics[width=0.9\textwidth]{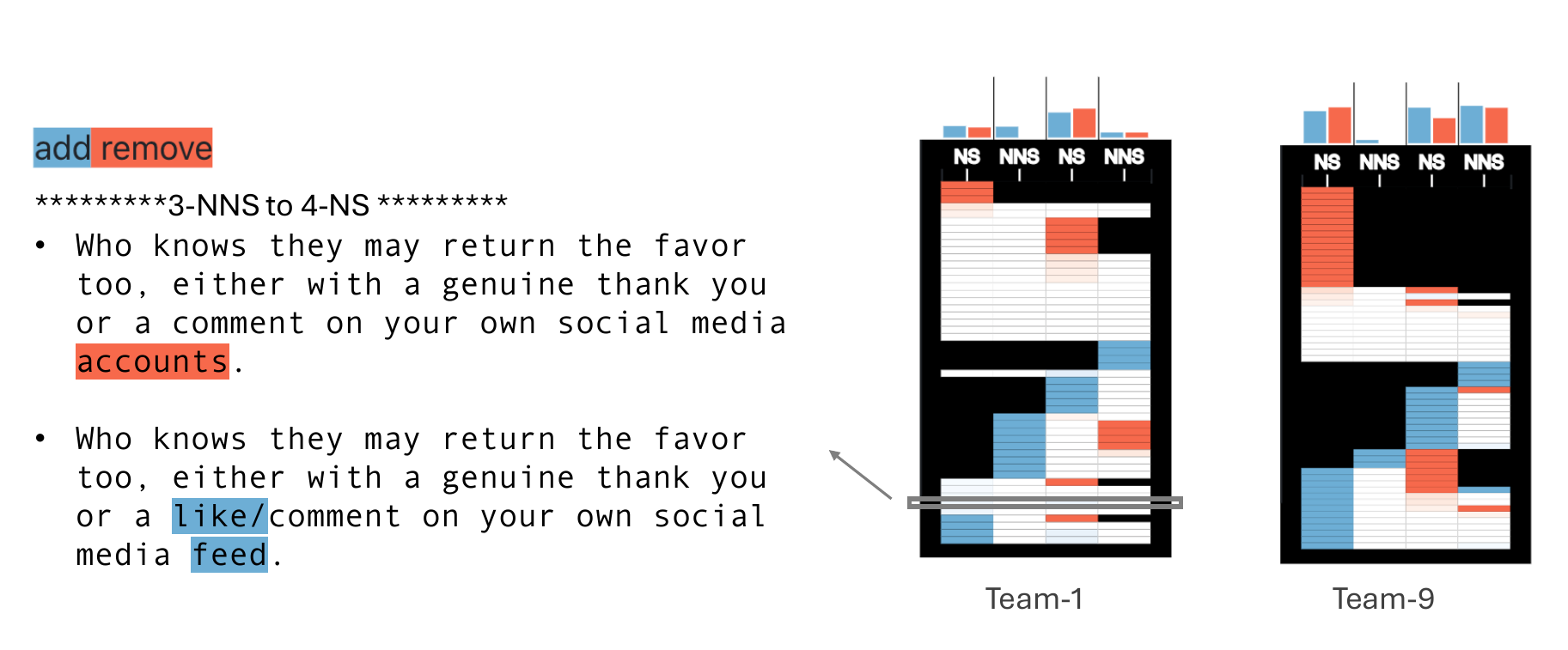}
    \caption{Sentence Flow of team-1 and team-9. Red: remove, blue: add, gray: edited by both authors.}
    \label{fig:sentence-flow}
\end{figure*}
\subsection{Understanding editing dynamics by branched time curves}

To tackle \textbf{T1}, we draw inspirations from the folded time curve proposed by Bach et al.~\cite{bach2015time}, where each dot represents a document version, the order on the curve denotes the temporal information, and the pairwise distance depends on the text similarity. We proposed a modified version as depicted in Figure~\ref{fig:branched-timecurve}: since authors write individually before collaborating on the same document, we adopted a two-head start, indicated by the gray arrows, which refer to the initial draft written by NS and NNS. Then, we use red and blue to refer to NNS and NS' contributions, respectively. The communication researchers suggest highlighting the differences between intermediate document versions $V_{inter}$ and end-of-the-day versions $V_{end}$, so we apply big filled dots for $V_{end}$ and small hollow dots for $V_{inter}$.

The branched time curves give an overview of the document's evolution, especially the ``V'' shape part, which shows how the merged version differs from NNS and NS' initial draft. Inspired by the proximity of the merged version to NS' version, in a recently published paper at a premier social computing conference (not cited here for anonymity), our collaborators conducted a statistical significance test and found that the merged document's lexical distance was indeed significantly closer to NS' initial writing. Meanwhile, in subsequent versions, NS' edits always result in a larger lexical distance than NNS. 

\subsection{Tracking specific content with Sentence Flow}
Though the time curves provide an overview on the document-level contributions from NS and NNS, the coarse granularity limits us from uncovering more content-specific insights (\textbf{T2}). Inspired by history flow~\cite{viegas2004studying, wang2015docuviz}, we propose a revised version called Sentence Flow, as depicted in Figure~\ref{fig:sentence-flow}. The x-axis is the author; the y-axis are individual sentences in the document. The edits of a sentence is color-coded: white for no activity, blue and red for word-level add and remove, and black for deleting the entire sentence. The darkness of red and blue indicates the edit distance, the darker the larger. On top of the sentence flow, there are also bar charts quantify the total edit distance. If users click on a sentence, they can see the content and change logs.

Our collaborators identified two dominant editing patterns in sentence flow. One is within-author editing, where authors primarily revise their own sentences and rarely modify others', e.g., in Team 1, adjacent colored areas are uncommon, with only one instance shown in Figure~\ref{fig:sentence-flow}, where the NS made minor modifications to NNS' sentences. The second pattern is between-author editing, as seen in Team 9 in Figure~\ref{fig:sentence-flow}, where several sentences were consecutively revised by different authors, resulting in a more balanced distribution.

Though the modified versions of existing visualizations reveal document modification behaviors of NS and NNS, it does not answer questions about how content are generated in collaborative writing. To address these questions, we conduct and present an event sequence analysis of content generation behaviors in the next section.

\end{document}